\documentclass[aps,pre,onecolumn,nofootinbib,superscriptaddress]{revtex4-2}

\usepackage{amsmath,amssymb}
\usepackage{times,hyperref,comment}
\hypersetup{
    colorlinks=true,       
    linkcolor=blue,          
    citecolor=blue,        
    filecolor=blue,      
    urlcolor=blue          
}
\usepackage[T1]{fontenc}

\renewcommand{\d}{{\text{d}}}
\newcommand{\T}{{\mathrm{T}}}
\newcommand{\grad}{{\boldsymbol{\nabla}}}
\newcommand{\acirc}{\overset{\alpha}{\circ}}
\newcommand{\lag}{\mathbb{L}}

\newcommand{\ie}{i.e., }
\newcommand{\eg}{e.g., }

\newcommand{\veca}{{\mathbf{a}}}
\newcommand{\vecb}{{\mathbf{b}}}
\newcommand{\vecD}{{\mathbf{D}}}
\newcommand{\vecF}{{\mathbf{F}}}
\newcommand{\vecf}{{\mathbf{f}}}
\newcommand{\vecH}{{\mathbf{H}}}
\newcommand{\vecI}{{\mathbf{I}}}
\newcommand{\vecJ}{{\mathbf{J}}}
\newcommand{\vecK}{{\mathbf{K}}}
\newcommand{\vecL}{{\mathbf{L}}}
\newcommand{\vecM}{{\mathbf{M}}}
\newcommand{\veco}{{\mathbf{0}}}
\newcommand{\vecp}{{\mathbf{p}}}
\newcommand{\vecq}{{\mathbf{q}}}

\newcommand{\vecr}{{\mathbf{r}}}
\newcommand{\vecW}{{\mathbf{W}}}
\newcommand{\vecw}{{\mathbf{w}}}

\newcommand{\vecX}{{\mathbf{X}}}
\newcommand{\vecx}{{\mathbf{x}}}
\newcommand{\vecY}{{\mathbf{Y}}}
\newcommand{\vecy}{{\mathbf{y}}}
\newcommand{\vecZ}{{\mathbf{Z}}}

\newcommand{\bmu}{{\boldsymbol{\mu}}}
\newcommand{\noise}{{\boldsymbol{\eta}}}
\newcommand{\error}{{\Delta\boldsymbol{\mathcal{E}}}}
\newcommand{\neverror}{{\Delta\mathcal{E}}}

\newcommand{\calI}{{\mathcal{I}}}
\newcommand{\calo}{{\mathcal{O}}}

\begin{document} 

\title{Consistent expansion of the Langevin propagator with application to entropy production}

\author{Benjamin Sorkin} 

\affiliation{School of Chemistry and Center for Physics and Chemistry of Living Systems, Tel Aviv University, 69978 Tel Aviv, Israel}



\author{Gil Ariel} 

\affiliation{Department of Mathematics, Bar-Ilan University, 52000 Ramat Gan, Israel}

\author{Tomer Markovich}

\email{tmarkovich@tauex.tau.ac.il}

\affiliation{School of Mechanical Engineering and Center for Physics and Chemistry of Living Systems, Tel Aviv University, 69978 Tel Aviv, Israel}

\begin{abstract}

Stochastic thermodynamics is a developing theory for systems out of thermal equilibrium. It allows to formulate a wealth of nontrivial relations among thermodynamic quantities such as heat dissipation, excess work, and entropy production in generic nonequilibrium stochastic processes. 
A key quantity for the derivation of these relations is the propagator\,---\,the probability to observe a transition from one point in phase space to another after a given time. 
Here, applying stochastic Taylor expansions, we devise a formal expansion procedure for the propagator of overdamped Langevin dynamics. The three leading orders are obtained explicitly. 
The technique resolves the shortcomings of the common mathematical machinery for the calculation of the propagator: For the evaluation of the first two displacement cumulants, the leading order Gaussian propagator is sufficient, however, some functionals of the propagator, such as the entropy production, which we refer to as ``first derivatives of the trajectory'', need to be evaluated to a previously-unrecognized higher order.
The method presented here can be extended to arbitrarily higher orders  to accurately compute any other functional of the propagator.

\end{abstract}

\maketitle

\tableofcontents

\section{Introduction}\label{sec:intro}

The theory of stochastic thermodynamics aims to find general relations among the statistics of thermodynamic quantities within microscopic (fluctuating) systems that are out of equilibrium. Examples include motor proteins, living systems, and swarming agents~\cite{book:stochastic}. 
It has seen many exciting developments in recent years. Notably, the Jarzynski relation~\cite{JarzynskiPRL97,JarzynskiPRE97} extends the second law of thermodynamics to such mesoscopic systems by establishing an equality between work statistics and the change in free energy in nonequilibrium transformations. Its extension, the Crooks relation~\cite{CrooksPRE99}, supplies additional information to the probability of work realizations. Theorems of this sort are often termed fluctuation relations, and could be unified by considering the statistics of the entropy production~\cite{SeifertRPP12,EspositoPRL10,HatanoPRL01}. The entropy production quantifies the extent to which a trajectory breaks time-reversal symmetry~\cite{SeifertRPP12,SeifertPRL05,book:stochastic} (or, colloquially, detailed balance). It has been shown that it coincides with the non-negative dissipated heat for vastly many scenarios~\cite{CatesEnt22,book:stochastic} (but excluding cases where the Einstein relation is invalid~\cite{SorkinPR24}). Thus, combined with trajectory-wise energy conservation proven by Sekimoto~\cite{sekimoto98Langevin_thermo}, one obtains the laws of classical thermodynamics by considering the stochastic trajectories in completely generic stochastic processes. 
Evidently, the estimation of entropy production and resolving the statistics of trajectories in nonequilibrium systems is a challenge worth resolving~\cite{KapplerPRX21,kapplerARX24,MartinJSTAT21,BiskerJSM2017,BiskerNATCOM19,GingrichPRL16,dunkelPRL21,CatesEnt22,fodor2016EP,SeifertPRL05,Parrondo07EPR,Parrondo08EPR,PRX2021}.

A key quantity for the resolution of path statistics is the propagator~\cite{LubenskyPRE07}\,---\,the probability distribution to obtain a particular displacement during some time duration. For Markov dynamics, it encodes the complete information of the system's evolution. For continuous-time systems (such as the overdamped Langevin equation considered here~\cite{book:schuss}), particularly important are the short-time expansions of the propagator due to the following reasons. (i) It suffices for the derivation of the Fokker-Planck equation corresponding to the Langevin equation based on its Kramers-Moyal expansion~\cite{book:FPE}. (ii) One is rarely capable of computing analytically or sampling experimentally the complete propagator for multidimensional systems with position- and time-dependent drift and diffusivity. Instead, one may utilize the available arsenal of short-time expansion methods~\cite{book:KloedenPlaten}.\footnote{In the opposite limit of late times, one may employ methods from large deviation theory~\cite{largdevPR09}.} (iii) In order to integrate properties over longer times, one may employ the path integral approach~\cite{book:schuss} and its various useful approximation methods inspired by quantum mechanics~\cite{LubenskyPRE07}, which consists of convolving many short-time propagators. Indeed, the Jarzynski equality~\cite{JarzynskiPRE97} and its generalizations (\eg the Hatano-Sasa relation~\cite{HatanoPRL01}) can be derived using path integral approaches. 

The aim of the present paper is to present a consistent expansion of the short-time propagator of the overdamped Langevin equation to arbitrary orders. We show that in many widely used applications, high order terms cannot be neglected, as they add up to observable, $\mathcal{O}(1)$, corrections over long $\mathcal{O}(1)$ time scales. The expansion is applied to obtain the entropy production rate in overdamped Langevin systems. The expansion resolves mathematical inconsistencies that are present in the current stochastic-thermodynamic literature~\cite{LubenskyPRE07,CatesEnt22,CugliandoloJPA2017}. Prior to delving into the derivation, we outline and summarize the main results.

\subsection{Plan and main results}\label{sec:plan}

The estimation of entropy production as well as the proof of fluctuation relations is facilitated by knowing the probability distribution to obtain a detailed trajectory. It is defined as follows. 
Consider the finely-discretized ($\Delta t \ll 0$) trajectory of a continuous-time stochastic system between times $0$ and $T=M\Delta t$, $\overrightarrow\vecx=\left(\vecx_0,\vecx_1,\ldots,\vecx_{M-1},\vecx_M\right)$, where $\vecx_i$ is the microstate of the system at time $i\Delta t$. The value of the forces (and other interventions into the system) is dictated by a time-varying {\em deterministic} protocol $\lambda (t)$, whose discrete sequence of values is $\overrightarrow\lambda=\left(\lambda_0,\lambda_1,\ldots,\lambda_{M-1},\lambda_M\right)$. We denote the probability to obtain the discretized trajectory $\overrightarrow\vecx$ under the effect of the protocol sequence $\overrightarrow\lambda$ as $\Pr[\overrightarrow\vecx,\overrightarrow\lambda]$. 

The probability to obtain a trajectory $\Pr[\overrightarrow\vecx,\overrightarrow\lambda]$ is a model-free quantity. Namely, it is defined and determined by the observed dynamics. In principle, one is capable of directly sampling it via experimental observations (usually, its marginals~\cite{BiskerJSM2017,dunkelPRL21,Parrondo08EPR}). Mathematically, it is a $M+1$-dimensional marginal for the continuous-time dynamics. In particular, it must not depend on how we choose to represent, approximate, or discretize the dynamics, which can be done in a number of equivalent ways. For example, in this paper we consider Langevin systems which we present in Sec.~\ref{sec:model}. Such processes can be modelled via a Fokker-Planck equation or a stochastic differential equation (SDE), which itself can be expressed using the It\^o or Stratanovich conventions~\cite{LubenskyPRE07,CPsokolov2010}; see details in Sec.~\ref{sec:dilemma}. These conventions are equivalent in the sense that any SDE written using the It\^o convention can be transferred into a different equation using the Stratanovich convention, and vice-versa, such that the two processes are the same (all finite dimensional marginals have the same distribution~\cite{book:oksendal}). The functional $\Pr[\overrightarrow\vecx,\overrightarrow\lambda]$ must not depend on the choice of convention.

To compute the entropy production, we aim to measure the breaking of time-reversal symmetry. Consider the time-reversed trajectory, $\overleftarrow\vecx=\left(\vecx_M,\vecx_{M-1},\ldots,\vecx_1,\vecx_0\right)$. We ask how likely is it to obtain this trajectory under the reversed protocol, $\overleftarrow\lambda=\left(\lambda_M,\lambda_{M-1},\ldots,\lambda_1,\lambda_0\right)$. Other than the reversal of the protocol $\overleftarrow\lambda$, the probabilistic laws of motion are the same: simply put\,---\,time still runs forward. Therefore, the probability for this reversed trajectory is given by the same functional above, but with the entries appearing in a reversed order, $\Pr[\overleftarrow\vecx,\overleftarrow\lambda]$; see Sec.~\ref{sec:EPR} for details.\footnote{Upon considering, say, the underdamped Langevin equation, phase space includes both positions and momenta, $\vecx=(\vecr,\vecp)$. Then, considering the reversed process, the momenta must invert sign: if $(\overrightarrow{\vecr},\overrightarrow{\vecp})$ describes the forward process, then $(\overleftarrow{\vecr},-\overleftarrow{\vecp})$ describes the reversed process. A similar parity issue appears when inverting magnetic forces, in which case the magnetic force due to protocol reversal must invert sign as well~\cite{OnsagerPR31}.}

The entropy production is a trajectory-dependent stochastic quantity measuring time-reversal symmetry breaking by the final time $T$, defined according to~\cite{book:stochastic,SeifertPRL05,CatesEnt22}
\begin{equation}
    \Sigma_T[\overrightarrow\vecx]=\ln\frac{\Pr[\overrightarrow\vecx,\overrightarrow\lambda]}{\Pr[\overleftarrow\vecx,\overleftarrow\lambda]},\label{eq:EPR}
\end{equation}
with $\Sigma_0=0$. The mean entropy production is
\begin{equation}
    \langle \Sigma_T \rangle=\int \d\overrightarrow\vecx\Pr[\overrightarrow\vecx,\overrightarrow\lambda] \ln\frac{\Pr[\overrightarrow\vecx,\overrightarrow\lambda]}{\Pr[\overleftarrow\vecx,\overleftarrow\lambda]},\label{eq:EPRav}
\end{equation}
where $\d\overrightarrow\vecx=\prod_{m=0}^M\d\vecx_m$. As a corollary, it is the Kullback-Liebler divergence~\cite{BOOK:infotheor} between the distribution to sample a trajectory $\overrightarrow\vecx$ (with protocol $\overrightarrow\lambda$) and the reference distribution to sample the reversed trajectory $\overleftarrow\vecx$ (with protocol $\overleftarrow\lambda$).
The average entropy production is therefore non-negative, where zero is only obtained for time-reversal-symmetric dynamics\,---\,those that are equally likely to undergo a particular path under a given forcing, and its reversed one. 

In this paper, we deal with diffusion processes that are orchestrated by the Langevin equation. Their path probabilities can be expressed via a Markov decomposition, $\Pr[\overrightarrow\vecx,\overrightarrow\lambda]=p(\vecx_0,0)\prod_{i=1}^M P(\vecx_i,\lambda_i|\vecx_{i-1},\lambda_{i-1})$. Note that any explicit time dependence on time is incorporated into the protocol $\lambda(t)$. In this expression, $p(\vecx,t)$ is the instantaneous probability to obtain a microstate $\vecx$ at time $t$ and $P(\vecx',\lambda'|\vecx,\lambda)$ is the propagator we wish to compute, evaluated at vanishingly-short times $\Delta t\ll 0$. The Markov property allows us to introduce the following separation of the entropy production {\em rate} at time instant $t=i\Delta t$,
\begin{equation}
    \dot\Sigma_t=\dot S_t+\dot\Omega_t,\qquad\dot\Omega_t=\lim_{\Delta t\to0}\frac{1}{\Delta t}\ln\frac{P(\vecx_{i}+\Delta\vecx_i,\lambda_{i}+\Delta\lambda_{i}|\vecx_{i},\lambda_{i})}{P(\vecx_{i},\lambda_{i}|\vecx_{i}+\Delta\vecx_i,\lambda_{i}+\Delta\lambda_{i})}\,,\label{eq:infoheat}
\end{equation}
where $S_t=-\ln p(\vecx_i,t)$ is the instantaneous stochastic Shannon entropy~\cite{SeifertPRL05}. We will henceforth refer to $\dot\Omega_t$ as the informatic heat rate (due to its coincidence with the heat in many scenarios mentioned above~\cite{CatesEnt22}). In Eq.~\eqref{eq:infoheat}, we defined $\Delta\vecx_i=\vecx_{i+1}-\vecx_i$ (which is $\sim\Delta t^{1/2}$ for diffusion processes), and $\Delta\lambda_i=\lambda_{i+1}-\lambda_i$ (which is of order $\Delta t$ since the protocol is deterministic and set externally).
The definition of the protocol is convenient to see the reversal of the forcing. Without loss of generality, for the rest of the derivation we take $\lambda_i=i\Delta t$.

Equation~\eqref{eq:infoheat} shows why the short-time propagator is the sought-after quantity. Clearly, it is required in order to compute the entropy production rate at a given instant, which then could be integrated over to find the accumulated entropy production during the entire continuous-time Markov process. Additionally, convolving many short-time propagators resolves the entire trajectory statistics. It is once again important to stress that the propagator is a physical quantity which in principle can be sampled directly from experimental or numerical data. Once it is known (analytically or from sampling), Eq.~\eqref{eq:infoheat}, in a sense, requires to extrapolate it to the short time limit. It, therefore, does not contain any discretization and must not depend on $\Delta t$ or the chosen approximation scheme. Since the propagator is analytically tractable only under rare circumstances (see Appendix~\ref{app:GBM} for one example, where we then extract the short-time limit from a known propagator for comparison), we have no choice but to find it from a consistent short-time expansion.

The main result of this paper is Eq.~\eqref{eq:propGEN}\,---\,a general expression for the propagator of Langevin equations, correct to arbitrary $\Delta t$. At the same time, it can directly be Taylor-expanded for small $\Delta t$; see details in Sec.~\ref{sec:propagator}. There, to order $\Delta t$, we arrive at an expansion which takes the form
\begin{eqnarray}
    P(\vecx+\Delta\vecx,\lambda+\Delta \lambda|\vecx,\lambda)&=&P_{1/2}(\vecx+\Delta\vecx,\lambda+\Delta \lambda|\vecx,\lambda)\nonumber\\
    &\times&\left[1+\Delta\vecx\cdot\boldsymbol\Phi\left(\left.\frac{\Delta\vecx\Delta\vecx}{\Delta t}\right|\vecx,\lambda\right)+\Delta t\Psi\left(\left.\frac{\Delta\vecx\Delta\vecx}{\Delta t}\right|\vecx,\lambda\right)+\calo(\Delta t^{3/2})\right]\,.\label{eq:propSTRUC}
\end{eqnarray}
Here, $P_{1/2}(\vecx+\Delta\vecx,\lambda+\Delta \lambda|\vecx,\lambda)$ is the leading-order Gaussian propagator of diffusive systems; Eq.~\eqref{eq:propSIMP} below. (The subscript $1/2$ will be explained in Sec.~\ref{sec:prop1/2}.) The expressions $\boldsymbol\Phi(\vecK|\vecx,\lambda)$ and $\Psi(\vecK|\vecx,\lambda)$ are (currently unspecified) functions, which we find to be polynomial in the rank-two tensor $\vecK = \Delta\vecx\Delta\vecx/\Delta t$ and are of order 1. We provide their explicit form in Eq.~\eqref{eq:phi} for $\boldsymbol\Phi$ and Eq.~\eqref{eq:psi} for $\Psi$. Clearly, the propagator is not Gaussian in general.

Typical dynamical averages, such as correlation functions or response functions, require only the leading-order propagator, $P_{1/2}(\vecx+\Delta\vecx,t+\Delta t|\vecx,t)$~\cite{LubenskyPRE07}. However, the entropy production [Eq.~\eqref{eq:infoheat}] is a ``first derivative of the trajectory''\,---\,a difference between two quantities that vary by $\Delta\vecx\sim\Delta t^{1/2}$ and $\Delta t$, divided by $\Delta t$. This division requires that we have the logarithm accurate upto and including order $\Delta t$. Specifically, inserting Eq.~\eqref{eq:propSTRUC} into Eq.~\eqref{eq:infoheat}, with the forward and reverse paths we have,
\begin{eqnarray}
    \dot\Omega_t&=&\lim_{\Delta t\to0}\frac{1}{\Delta t}\ln\frac{P_{1/2}(\vecx_{i}+\Delta\vecx_i,\lambda_{i}+\Delta\lambda_{i}|\vecx_{i},\lambda_{i})}{P_{1/2}(\vecx_{i},\lambda_{i}|\vecx_{i}+\Delta\vecx_i,\lambda_{i}+\Delta\lambda_{i})}\nonumber\\
    &&+\frac{\Delta\vecx_i}{\Delta t}\cdot\left[\boldsymbol\Phi\left(\left.\frac{\Delta\vecx_i\Delta\vecx_i}{\Delta t}\right|\vecx_i,\lambda_i\right)+\boldsymbol\Phi\left(\left.\frac{\Delta\vecx_i\Delta\vecx_i}{\Delta t}\right|\vecx_i+\Delta\vecx_i,\lambda_i+\Delta \lambda_i\right)\right]\nonumber\\
    &&-\frac{\Delta\vecx_i\Delta\vecx_i}{2\Delta t}:\left[\boldsymbol\Phi\left(\left.\frac{\Delta\vecx_i\Delta\vecx_i}{\Delta t}\right|\vecx_i,\lambda_i\right)\boldsymbol\Phi\left(\left.\frac{\Delta\vecx_i\Delta\vecx_i}{\Delta t}\right|\vecx_i,\lambda_i\right)\right.
    \nonumber\\&&\qquad\qquad\qquad\qquad\qquad\left.-\boldsymbol\Phi\left(\left.\frac{\Delta\vecx_i\Delta\vecx_i}{\Delta t}\right|\vecx_i+\Delta\vecx_i,\lambda_i+\Delta\lambda_i\right)\boldsymbol\Phi\left(\left.\frac{\Delta\vecx_i\Delta\vecx_i}{\Delta t}\right|\vecx_i+\Delta\vecx_i,\lambda_i+\Delta\lambda_i\right)\right]\nonumber\\
    &&+\left[\Psi\left(\left.\frac{\Delta\vecx_i\Delta\vecx_i}{\Delta t}\right|\vecx_i,\lambda_i\right)-\Psi\left(\left.\frac{\Delta\vecx_i\Delta\vecx_i}{\Delta t}\right|\vecx_i+\Delta\vecx_i,\lambda_i+\Delta\lambda_i\right)\right]+\calo(\Delta t^{1/2})\,,\label{eq:insertinfoheat}
\end{eqnarray}
where we used the shorthand notation for the inner products $\grad\cdot\veca=\nabla^\mu a^\mu$ and $\grad\grad:\vecD=\nabla^\mu\nabla^\nu D^{\mu\nu}$, and the Einstein notation is implied.\footnote{To avoid ambiguity, the double-dot notation will only be used in quadratic forms.}  First, the leading order propagators in each direction are equal to leading order, $P_{1/2}(\vecx_{i}+\Delta\vecx_i,\lambda_{i}+\Delta\lambda_{i}|\vecx_{i},\lambda_{i})/P_{1/2}(\vecx_{i},\lambda_{i}|\vecx_{i}+\Delta\vecx_i,\lambda_{i}+\Delta\lambda_{i})=1 +\calo(\Delta t^{1/2})$. Again, this implies that division by $\Delta t$ requires including higher order difference between them. The corrections $\Delta\vecx\cdot\boldsymbol\Phi$ and $\Delta t\Psi$, appearing from higher-order integration methods of the Langevin equation are respectively of order $\Delta t^{1/2}$ and $\Delta t$, and hence cannot be neglected \emph{a priori}\,---\,they carry order-$1$ [last three rows of Eq.~\eqref{eq:insertinfoheat}] and even order-$\Delta t^{-1/2}$ [second row of Eq.~\eqref{eq:insertinfoheat}] contributions to the entropy production. This also shows why we may stop at the first three leading orders\,---\,the rest bear contributions of order $\Delta t^{1/2}$, which would vanish as $\Delta t \to 0$.

Previous approaches~\cite{LubenskyPRE07,CugliandoloJPA2017,CatesEnt22,CugliandoloREV23,CugliandoloJPA19,OnsagerPRB53} have approximated the short-time propagator for Langevin processes using a convention-dependent representation of the leading-order $P_{1/2}(\vecx+\Delta\vecx,\lambda+\Delta \lambda|\vecx,\lambda)$, on which we elaborate in Sec.~\ref{app:equiv}. We show that, in general, these approximations are missing the sought-after higher-order terms. The expansion scheme we derive for the propagator carries several consequences and advantages in this regard:
\begin{itemize}
    \item With the propagator in hand, it is natural to consider the first two cumulants of the displacement $\Delta \vecx$~\cite{LubenskyPRE07}. We show that these moments, found from any of the above-mentioned representations of the leading-order propagator are all correct to the same leading order, $\Delta t$. Thus, higher-order expansions carry no advantage with respect to the simplest (It\^o) representation of the leading-order propagator, and they all converge to the exact expression as $\Delta t \to 0$. 
    \item Previous derivations of the entropy production (see, for example, Ref.~\cite{CatesEnt22}) again used only the leading-order Gaussian propagator $P_{1/2}$ and assumed specific conventions for the forward and backward trajectories. As discussed above, this is not consistent with the understanding that both the propagator and the entropy production, Eq.~\eqref{eq:infoheat}, are discretization-free quantities. Below, we show that their result is correct due to a surprising cancellation of wrongfully-neglected terms. This cancellation is due to symmetries that are specific to the entropy production, as we elaborate in Sec.~\ref{sec:toy}. 
    \item For all functionals that are first derivatives of trajectories (such as the entropy production), \ie functionals consisting of a difference among two path-dependent quantities that only differ by $\Delta \vecx$ and $\Delta t$, divided by $\Delta t$, the order-$\Delta t$ term in \eqref{eq:propSTRUC} vanishes \textit{a posteriori} in the limit $\Delta t \to 0$. As a result, only the next-order corrections to the leading-order propagator, $\Delta \vecx\cdot\boldsymbol\Phi$, must be considered. For the same reasons, ``second derivatives'' of trajectories will require this term again, along with additional higher-order corrections to the propagator.
    \item To illustrate the importance of consistent propagator expansions, in Sec.~\ref{sec:toy} we suggest toy examples which are first derivatives of the trajectory, in which the convention applied in, e.g.,  Ref.~\cite{CatesEnt22}, may yield erroneous results. 
\end{itemize}

A few similar efforts to find consistent short-time expansions of the diffusive propagator beyond the leading order can be found in the literature. In parallel to us, Ref.~\cite{kapplerARX24} presented a perturbation theory approach to calculate the short-time propagator for the one-dimensional Fokker-Planck equation. In fact, Ref.~\cite{kapplerARX24} show expansion of higher order than presented here. As we explain below, we comment that the Langevin equation in one dimension is simpler for treatment not only because of the absence of Einstein-notation indices, but also since the first correction can be reproduced from a leading-order expansion performed specifically in the Stratonovich convention. Refs.~\cite{AitSahaliaECON02,AitSahaliaAT08} perform a direct expansion of the Fokker-Planck propagator in terms of the mutlivariate Hermite polynomials. Since the Fokker-Planck equation is equivalent to the Langevin equation considered here, our expansions\,---\,based on stochastic-Taylor expansions of the propagator's characteristic function\,---\,must coincide with theirs. By considering the Langevin equation, we are able to directly compare and contrast our results with the stochastic-thermodynamic literature. This facilitates a better understanding of potential sources of error and a comprehensive discussion on the applicability and usefulness of different discretization conventions. References~\cite{DrozdovPRE97,DrozdovJCP97} introduced the first correction to the leading-order Gaussian propagator by considering the propagator's moment generating function. Here, we do not truncate the expansion at an early stage as the latter do, and hence, in principle, we may go up to arbitrarily higher orders.

\subsection{Paper outline}

The paper is organized as follows. In Sec.~\ref{sec:model}, we present our model dynamics\,---\,the overdamped Langevin equation of motion corresponding to a given Fokker-Planck equation. We explain how it is constructed~\cite{book:FPE,book:schuss}, recall how to resolve the It\^o dilemma~\cite{book:FPE,LubenskyPRE07} (there is, in fact, no dilemma), and introduce our notation. In Sec.~\ref{sec:expans}, we recall how to consistently find short-time expansions beyond the leading-order~\cite{book:KloedenPlaten}. Section~\ref{sec:propagator} is the core of the paper, in which we present a consistent expansion scheme for the short-time propagator in overdamped Langevin systems which directly shows what order it is correct to. We do so by writing its Fourier-transform, employ the introduced stochastic Taylor expansions, and carry out the Hubbard-Stratonovich transformation~\cite{HubbardPRL59,StratonovichDAN57}. In Sec.~\ref{sec:EPR}, we show a consistent calculation for the entropy production. We make an extensive comparison between the present scheme and the shortcomings of previous methods, including
the insufficiently-high order calculation of the entropy production in Sec.~\ref{app:equiv}. We conclude with a summary of our findings and perspectives in Sec.~\ref{sec:summary}. Appendix~\ref{app:GBM} shows the coincidence of our results with expansions of an exactly solvable, one-dimensional model - the geometric Brownian motion.

\section{Model}\label{sec:model}

The aim of this section is two-fold. First, to connect the commonly-found physical notations and intuition regarding diffusion processes to the ones in the literature on SDEs. Second, to resolve the It\^o dilemma prior to proceeding deeper into the derivation. 
This is only presented for completeness and review purposes. The textbook by Schuss~\cite{book:schuss} presents a comprehensive analysis of the mathematical aspects underlying stochastic processes in physics.

\subsection{Constructing the Fokker-Planck equation}\label{sec:FPE}

We consider a system which is characterized by $N$ arbitrary random variables (degrees of freedom) at time $t$, denoted as $\vecX_t=(X_t^1,X_t^2,\ldots,X_t^N)$. For example, these may be the positions of $N/d$ particles in $d$ dimensions. We assume that these degrees of freedom undergo some Markovian diffusion process. Thus, the probability density function $\Pr(\vecX_t=\vecx)\equiv p(\vecx,t)$ follows a Fokker-Planck equation~\cite{book:FPE}
\begin{equation}
    \frac{\partial p(\vecx,t)}{\partial t}=-\grad\cdot[\veca(\vecx,t)p(\vecx,t)]+\grad\grad:[\vecD(\vecx,t)p(\vecx,t)]\,,\label{eq:FPE}
\end{equation}
where $\veca$ is the drift and $\vecD$ is the diffusivity. A given system is characterized by unique $\veca$ and $\vecD$ which must be specified \textit{a priori}.
We may write Eq.~\eqref{eq:FPE} as a probability conservation equation by identifying the probability flux $\vecJ(\vecx,t)$,
\begin{equation}
    \frac{\partial p(\vecx,t)}{\partial t}=-\grad\cdot\vecJ(\vecx,t),\qquad\vecJ(\vecx,t)=\veca(\vecx,t) p(\vecx,t)-\grad\cdot[\vecD(\vecx,t) p(\vecx,t)]\,.\label{eq:FPEflux}
\end{equation}
The expansions of Refs.~\cite{kapplerARX24,AitSahaliaECON02,AitSahaliaAT08} are based on this equation.

The use of Eq.~\eqref{eq:FPE} is abundant in physics. For example, if one is given with a particle that is immersed in a very viscous fluid [so the response of the particle to an external force $\vecf(\vecx,t)$ is linear via some mobility tensor, $\bmu(\vecx,t)$], and the particle undergoes diffusion due to random thermal fluctuations [quantified by a diffusion tensor $\vecD(\vecx,t)$], the probability distribution function to find the particle at $\vecx$, $p(\vecx,t)$, is governed by the Smoluchowski equation
\begin{equation}
    \frac{\partial p(\vecx,t)}{\partial t}=-\grad\cdot[\bmu(\vecx,t)\cdot\vecf(\vecx,t)p(\vecx,t)]+\grad\cdot[\vecD(\vecx,t)\cdot\grad p(\vecx,t)]\,.\label{eq:diffeq3}
\end{equation}
The drift here, therefore, is given by $\veca=\bmu\cdot\vecf+\grad\cdot\vecD$. The second term is called the spurious drift, and it ensures that if (1) the forces are conservative, $\vecf(\vecx)=-\grad H(\vecx)$, and (2) there mobility and diffusivity are connected via a uniform (inverse) temperature scalar, $\bmu(\vecx,t)=\beta\vecD(\vecx,t)$, then the Boltzmann factor $p(\vecx,t\to\infty)\sim e^{-\beta H(\vecx)}$ is attained at steady state.

Note that a particular system will follow a single Fokker-Planck equation, with specified drift $\veca$ and diffusivity $\vecD$ that can be constructed in a similar manner to the above. In the context of the upcoming It\^o dilemma, it is key to understand that there is no ambiguity as to the drift one has in a given scenario. Likewise (under some sufficient technical assumptions~\cite{book:schuss,book:oksendal}, which in this manuscript will always be assumed to hold), there exists a unique solution $p(\vecx,t)$ of Eq.~\eqref{eq:FPE} under a given initial condition $p(\vecx,0)$. As a consequence, the propagator $\Pr(\vecX_{t+\Delta t}=\vecx+\Delta\vecx|\vecX_t=\vecx)\equiv P(\vecx+\Delta\vecx,t+\Delta t|\vecx,t)$ of Eq.~\eqref{eq:FPE} [with initial conditions $P(\vecx+\Delta\vecx,t|\vecx,t)=\delta(\Delta\vecx)$] also exists and is unique.

\subsection{Constructing the Langevin equation}\label{sec:langevin}

There are a wealth of tools allowing to solve the Fokker-Planck equation~\cite{book:FPE}. For example, many operator-identities motivated by quantum mechanics provide the propagator operator of a given equation. Another approach is the Feynmann-Kac formula~\cite{KacAMS49,book:FPE,book:schuss}, expressing expectation values via a path integral. As mentioned in Sec.~\ref{sec:intro}, the path integral approach has proven useful in stochastic thermodynamics. We will draw motivation from this approach, and resolve the transition statistics at short times. For that purpose, we construct the Langevin equation. Instead of writing the partial differential equation for the microstates instantaneous probability density function, the Langevin equation is a SDE for the random evolution of the microstate itself.

Using either the Kramers-Moyal expansion or the aforementioned path-integral formulation, it is possible to show that  the stochastic integral equation corresponding to the Fokker-Planck equation [Eq.~\eqref{eq:FPE}] is
\begin{equation}
    \vecX_{t+\Delta t}-\vecX_t=\int_t^{t+\Delta t}\veca(\vecX_{t'},t')\d t'+\int_t^{t+\Delta t}\vecb(\vecX_{t'},t')\cdot\d\vecW_{t'}\,.\label{eq:langevin}
\end{equation}
In this expression, $\vecb$ is the noise amplitude matrix, obeying $\vecD=\vecb\cdot\vecb^\T/2$. Additionally, $\vecW_t$ is a Wiener process, meaning that the increments $\vecW_{t'}-\vecW_{t}$ ($t'>t$) are normal distributed with variance tensor $(t'-t)\vecI$ and zero mean, and are independent of past realizations~\cite{book:schuss}. The dot product among $\vecb$ and $\d\vecW_{t'}$ is interpreted in the It\^o convention; see Sec.~\ref{sec:dilemma} below. 

The integral Eq.~\eqref{eq:langevin} is commonly abbreviated as the SDE
\begin{equation}
    \d\vecX_t=\veca(\vecX_t,t)\d t+\vecb(\vecX_t,t)\cdot\d\vecW_t\,.\label{eq:SDE}
\end{equation}
This is not a numerical approximation, rather, a differential notation for an integral expression. Thus, we have not chosen a discretization of $\d t\to\Delta t$ or approximated the integrals via the Euler-Maruyama method (see Sec.~\ref{sec:expans} below); it is a shorthand, exact notation. In order to obtain short time-scales expansion ($\Delta t\to0$), one should properly Taylor expand Eq.~\eqref{eq:langevin}; this is done in Sec.~\ref{sec:expans} below. 

One typically finds the following notation in the physical literature for the Langevin equation~\cite{book:schuss},
\begin{equation}
    \dot\vecX_t=\veca(\vecX_t,t)+\noise(\vecX_t,t)\,,\label{eq:langPhys}
\end{equation}
where $\noise$ is a Gaussian white noise, having variance $\langle\noise(\vecX_{t'},t')\noise(\vecX_t,t)|\vecX_t\rangle=2\vecD(\vecX_t,t)\delta(t'-t)$ ($t'>t$). It is equivalent to Eqs.~\eqref{eq:langevin} and~\eqref{eq:SDE} upon replacing the noise as $\noise(\vecx,t)\d t=\vecb(\vecx,t)\cdot\d\vecW_t$. However, one must specify the noise interpretation so Eq.~\eqref{eq:langPhys} will be well-defined; see Sec.~\ref{sec:dilemma}. Otherwise, the physical notation Eq.~\eqref{eq:langPhys} is ambiguous, except in the simple case in which $D$ does not depend on $\vecx$.

The drift $\veca(\vecx,t)$ appearing in these stochastic equations is exactly as in Sec.~\ref{sec:FPE} so as to reproduce the correct Fokker-Planck equation (also called the Kolmogorov forward equation associated with the SDE~\cite{book:oksendal}). For example, the Langevin equation describing the motion of the particle under the external force $\vecf(\vecx,t)$ in a viscous medium with mobility $\bmu(\vecx,t)$ and diffusivity $\vecD(\vecx,t)$ reads
\begin{equation}
    \d\vecX_t=[\bmu(\vecX_t,t)\cdot\vecf(\vecX_t,t)+\grad\cdot\vecD(\vecX_t,t)]\d t+\vecb(\vecX_t,t)\cdot\d\vecW_t\,.\label{eq:SDE3}
\end{equation}
In this expression, a tensor diffusivity means that one should find a noise amplitude $\vecb$ such that $\vecD=\vecb\cdot\vecb^T/2$. The point we wish to emphasize here is that there is a unique stochastic differential equation corresponding to the unique Fokker-Planck equation describing the evolution of a system. In the example seen here, Eq.~\eqref{eq:SDE3} specifically gives rise to Eq.~\eqref{eq:diffeq3}, and in general Eq.~\eqref{eq:SDE}\,---\,to Eq.~\eqref{eq:FPE}.

\subsection{The It\^o dilemma}\label{sec:dilemma}

We recall that the second term in Eq.~\eqref{eq:langevin} is interpreted in the It\^o convention. The meaning of this can be understood through discretization. Denoting the discrete time step $\delta t$, the first integral in Eq.~\eqref{eq:langevin} can be simply interpreted as a Riemann sum. However, the definition of the second integral requires careful consideration. Since $\d\vecW_{t'}\sim\d t^{1/2}$, the position $\vecX_{t'}$ at which $\vecb$ is evaluated is not arbitrary. Namely, an integral involving the It\^o convention [as in Eq.~\eqref{eq:langevin}] is finely-discretized as~\cite{book:schuss}
\begin{equation}
    \boldsymbol\calI_0\equiv\int_{t}^{t+\Delta t}\vecb(\vecX_{t'},t')\cdot\d\vecW_{t'}=\lim_{\delta t\to0}\sum_{i=1}^{\Delta t/\delta t}\vecb(\vecX_{t+(i-1)\delta t},t+(i-1)\delta t)\cdot[\vecW_{t+i\delta t}-\vecW_{t+(i-1)\delta t}]\,,\label{eq:disc_Ito_product}
\end{equation}
\ie the function $\vecb$ is specifically evaluated at the starting point of each interval, which leads to a Markovian process. 

To evaluate $\vecb$ at some other intermediate point between $(i-1)\delta t$ and $i\delta t$, $\vecX_{t+(i+\alpha-1)\delta t}=\alpha \vecX_{t+i\delta t}+(1-\alpha) \vecX_{t+(i-1)\delta t}$, the corresponding integral reads instead~\cite{book:schuss}
\begin{equation}
    \boldsymbol\calI_\alpha\equiv\int_{t}^{t+\Delta t}\vecb(\vecX_{t'},t')\acirc\d\vecW_{t'}=\lim_{\delta t\to0}\sum_{i=1}^{\Delta t/\delta t}\vecb(\vecX_{t+(i+\alpha-1)\delta t},t+(i-1)\delta t)\cdot[\vecW_{t+i\delta t}-\vecW_{t+(i-1)\delta t}]\,.
\end{equation}
In order to compare this with Eq.~\eqref{eq:disc_Ito_product}, we Taylor-expand $\vecb$ around $\vecX_t$ for small $\delta\vecX_t=\vecX_{i\delta t}-\vecX_{(i-1)\delta t}$. Since there are $\sim1/\delta t$ summands, we should compute each summand to order $\delta t$ and neglect corrections of order $\delta t^{3/2}$. Combined with Eq.~\eqref{eq:langevin}, this yields a correction relative to $\boldsymbol\calI_0$ that does not vanish as $\delta t \to 0$,
\begin{eqnarray}
    \calI^\mu_\alpha
    &=&\lim_{\delta t\to0}\sum_{i=1}^{\Delta t/\delta t}b^{\mu\nu}\bigg( \left[ \vecX_{t+(i-1)\delta t},t+(i-1)\delta t \right] \left[ W^\nu_{t+i\delta t}-W^\nu_{t+(i-1)\delta t} \right]\nonumber\\
    &+&\alpha b^{\sigma\rho} \left( \nabla^\sigma b^{\mu\nu}\right) \left[\vecX_{t+(i-1)\delta t},t+(i-1)\delta t \right] \left[ W^\nu_{t+i\delta t}-W^\nu_{t+(i-1)\delta t} \right] \left[W^\rho_{t+i\delta t}-W^\rho_{t+(i-1)\delta t} \right] \bigg)\nonumber\\
    &=&\calI^\mu_0+\lim_{\delta t\to0}\sum_{i=1}^{\Delta t/\delta t}\alpha b^{\sigma\nu}(\nabla^\sigma b^{\mu\nu}) \left[ \vecX_{t+(i-1)\delta t},t+(i-1)\delta t \right]\delta t\,,\label{eq:bdW_alpha}
\end{eqnarray}
where we used $\left[ W^\nu_{t+i\delta t}-W^\nu_{t+(i-1)\delta t} \right] \left[W^\rho_{t+i\delta t}-W^\rho_{t+(i-1)\delta t} \right] =\delta^{\nu\rho}\delta t+\calo(\delta t^{3/2})$. 
Evaluating $\vecb$ also at $t+(i+\alpha-1)\delta t$ gives corrections relative to $\boldsymbol\calI_0$ that are of order $\delta t^{3/2}$ and hence negligible.

With that, we recapped the root of the It\^o dilemma. Namely, given a stochastic equation with position-dependent diffusivity [\eg Eq.~\eqref{eq:langevin}], without specifying the convention $\alpha$ adopted within the product acting among $\vecb$ and the differential increment $\d\vecW_{t'}$, the equation is ambiguous. 
The case $\alpha=0$ is called the It\^o convention, $\alpha=1/2$ is called the Stratonovich convention, and $\alpha=1$ is often called the H\"anggi convention~\cite{LubenskyPRE07,CPsokolov2010}. For brevity, we will use the usual conventions for denoting the It\^o and Stratonovich integrals, $\vecb\overset{0}{\circ}\d\vecW=\vecb\cdot\d\vecW$ and $\vecb\overset{1/2}{\circ}\d\vecW=\vecb\circ\d\vecW$.

Now, the question is, therefore, how to reproduce the right Fokker-Planck equation. Through Eq.~\eqref{eq:bdW_alpha}, we see that Eq.~\eqref{eq:langevin} can be rewritten in a different convention as
\begin{equation}
    \vecX_{t+\Delta t}-\vecX_t=\int_t^{t+\Delta t}\veca_\alpha(\vecX_{t'},t')\d t'+\int_t^{t+\Delta t}\vecb(\vecX_{t'},t')\acirc\d\vecW_{t'}\,,\label{eq:langevin_alpha}
\end{equation}
where $a^\mu_\alpha=a^\mu_0-\alpha b^{\sigma\nu}\nabla^\sigma b^{\mu\nu}$ and for brevity we will denote $\veca_0=\veca$. In other words, there are infinitely many representations, characterized by $\alpha\in[0,1]$, which are all equivalent as long as one introduces the right convention-dependent drift term~\cite{LubenskyPRE07,CatesEnt22}. This implies that they describe the \emph{same} stochastic process (in the sense that all finite dimensional marginals are the same~\cite{book:oksendal}). In particular, they all correspond to the \emph{same} Fokker-Planck equation [Eq.~\eqref{eq:FPE}]. Therefore, the propagator must be identical for all conventions. The latter realization has not been the consensus so far~\cite{LubenskyPRE07,CugliandoloJPA2017,CatesEnt22,CugliandoloREV23,CugliandoloJPA19,OnsagerPRB53}, and the aim of the paper is to correct the mathematical shortcomings that ensue.

To summarize, although there appear to be infinitely many equations depending on the convention $\alpha$, the drift must be constructed such that all equations would give rise to the \emph{same} stochastic process. 
Thus, there is no dilemma\,---\,once one has picked a convention for the product, the appropriate drift must be included so to get the correct Fokker-Planck equation. Most importantly, if one observes a dependence on the discretization scheme, it implies that some terms were not included.

\section{Stochastic Taylor expansions}\label{sec:expans}

In this section we recall the general strategy of stochastic Taylor expansions. The methods applied throughout this section are outlined in the textbook by Kloeden and Platen.~\cite{book:KloedenPlaten}. Note that the focus of Ref.~\cite{book:KloedenPlaten} is the construction of high-order numerical integration schemes for the a given SDE. In contrast, our goal is to write analytic approximations to the propagator of the exact SDE. The two goals are not too different, however, as we will show below, the former requires including an extra term which for our purposes will be negligible.

The starting point is the integral formulation for the displacement, Eq.~\eqref{eq:langevin}, which is exact. This formulation shows that the full distribution of $\Delta\vecX_t=\vecX_{t+\Delta t}-\vecX_t$, i.e., to all orders in $\Delta t$, depends on the entire Wiener process $\{\vecW_{s}\}_{s=t}^{t+\Delta t}$. Note that accoring to the Markov property, the entire past before time $t$ is summarised in the initial condition $\vecX_t$.
Below, the expansions are applied to write the displacements, as a function of noise realizations, up to high orders in $\Delta t$, which is assumed to be small. The derivation will be carried with the It\^o convention for the product; the equivalence among conventions at this level of stochastic expansions is discussed in Sec.~\ref{app:equiv}.
 
The expansion consists of the following steps:
\begin{enumerate}
    \item Write the exact displacement $\vecX_s=\vecX_t+\int_t^s\d\vecX_{s'}$ for $t\leq s\leq t+\Delta t$ starting from the the initial point $s=t$.
    \item Expand every function $g(\vecX_s,s)$ which may appear within the integrals of Eq.~\eqref{eq:langevin} in a Taylor series. Remembering that $\d\vecX_t\sim\d t^{1/2}$. For example, up to order $\Delta t$ we get:
    \begin{equation}
        g(\vecX_s,s)=g(\vecX_t,t)+\left[\grad g(\vecX_t,t)\right]\cdot\int_t^s\d\vecX_{s'}+\frac{1}{2}\left[\grad\grad g(\vecX_t,t)\right]:\int_t^s\d\vecX_{s'}\int_t^s\d\vecX_{s''}+\frac{\partial g(\vecX_t,t)}{\partial t}\int_t^s\d s'+\mathcal{O}(\Delta t^{3/2})\,.\label{eq:integrand_exp1}
    \end{equation}
    \item Apply the It\^o lemma: wherever $\d\vecX_t$ appears, we insert Eq.~\eqref{eq:SDE}. For example, the second term in Eq.~\eqref{eq:integrand_exp1} contains
    \begin{equation}
        \int_t^s\d\vecX_{s'}=\int_t^s\veca(\vecX_{s'},s')\d s'+\int_t^s\vecb(\vecX_{s'},s')\cdot\d\vecW_{s'}\,.
    \end{equation}
    Once again the integrands contain a function which we should be expanded recursively in small $\vecX_{s'}-\vecX_t$ and $s'-t$.
\end{enumerate}
This prescription is repeated until the desired order of convergence for $\Delta\vecX_t$ is achieved. We now present the first three orders, which will play a role later in the text.

\subsection{The Euler-Maruyama method}\label{sec:EulerMaruyamanMethod}

Trivially, the lowest-order expansion, being of order-$\Delta t^{1/2}$~\cite{book:KloedenPlaten}, is obtained by keeping the leading order terms from the integrals of Eq.~\eqref{eq:langevin},
\begin{equation}
    \Delta \vecX_{t}=\veca\Delta t + \vecb\cdot\Delta \vecW_{t}+\calo(\Delta t)\,,\label{eq:dX1/2}
\end{equation}
where $\Delta\vecW_t=\vecW_{t+\Delta t}-\vecW_t$ and, henceforth, unless specified, the expansion coefficients are evaluated at the initial point $(\vecX_{t},t)$. Equation~\eqref{eq:dX1/2} is called the Euler-Maruyama method for numerical integration of SDEs. 

The current literature on stochastic thermodynamics, \eg Refs.~\cite{LubenskyPRE07,CugliandoloJPA2017,CatesEnt22,CugliandoloREV23,CugliandoloJPA19,OnsagerPRB53}, have regarded Eq.~\eqref{eq:SDE} already as the discretized equation (With $\d t\to\Delta t$). As we briefly show in Sec.~\ref{sec:propSIMP}, their propagator has been derived from this expression. While Eq.~\eqref{eq:dX1/2} agrees with Eq.~\eqref{eq:langevin} up to and including order $\Delta t^{1/2}$, quantities requiring higher-orders estimated with either will not agree. The entropy production is one such quantity; see Sec.~\ref{sec:EPR}.

The first term of Eq.~\eqref{eq:dX1/2}, $\veca\Delta t$, cannot be discarded if one wants the numerical integration to produce the drift to the correct order~\cite{book:KloedenPlaten}. Removing it will result in a numerical SDE approximation scheme that has global order-$0$, i.e., the accumulated error after taking $\sim1/\Delta t$ steps will be $\calo(1)$. In other words, Ref.~\cite{book:KloedenPlaten} ``overshoots'' and keeps an order-$\Delta t$ term that cannot be approximated to fractional orders for correct convergence to order-$\Delta t^{1/2}$. Since we are not after a consistently-converging approximation of the SDE, but rather after a high-order Taylor expansion of the propagator, which we use to evaluate the entropy production, we will discard the analogous ``overshoot'' term in the order-$\Delta t^{3/2}$ approximation below. (We keep the one of Eq.~\eqref{eq:dX1/2} since it is low-order.)

\subsection{The Milstein method}

Going forward, we briefly demonstrate the above procedure and derive the order-$\Delta t$ stochastic expansion~\cite{book:KloedenPlaten}. The first term of Eq.~\eqref{eq:langevin} is already of order $\Delta t$ so, as before
\begin{equation}
    \int_t^{t+\Delta t}\veca(\vecX_{t'},t')\d t'=\veca(\vecX_t,t)\Delta t+\calo(\Delta t^{3/2})\,.
\end{equation}
Now, the second term of Eq.~\eqref{eq:langevin} is of leading-order $\Delta t^{1/2}$, hence we should keep the next-order term. We write 
\begin{eqnarray}
    \vecb(\vecX_{t'},t')&=&\vecb(\vecX_t,t)+\left[\nabla^\rho\vecb(\vecX_t,t)\right]\int_t^{t'}\d X_{t''}^\rho+\calo(\Delta t)\nonumber\\
    &=&\vecb(\vecX_t,t)+\left[\nabla^\rho\vecb(\vecX_t,t)\right]\int_t^{t'}b^{\rho\sigma}(\vecX_{t''},t'')\cdot\d W_{t''}^\sigma+\calo(\Delta t)\nonumber\\
    &=&\vecb(\vecX_t,t)+\left[b^{\rho\sigma}(\vecX_{t},t)\nabla^\rho\vecb(\vecX_t,t)\right]\int_t^{t'}\d W_{t''}^\sigma+\calo(\Delta t)\,.
\end{eqnarray}
Inserting this result in Eq.~\eqref{eq:langevin}, one finds the order-$\Delta t$ expansion,
\begin{equation}
    \Delta X^\mu_t=a^\mu\Delta t + b^{\mu\nu}\Delta W^\nu_{t}+F^{\mu\nu\sigma}\Delta Y^{\nu\sigma}_t +\calo(\Delta t^{3/2})\,,\label{eq:dX1}
\end{equation}
where, for brevity, we defined $F^{\mu\nu\sigma}=b^{\rho\sigma}\nabla^\rho b^{\mu\nu}$ and $\Delta Y^{\nu\sigma}_t = \int_{t}^{t+\Delta t}\d W^\nu_{t'}\cdot\int_{t}^{t'}\d W^\sigma_{t''}$; the latter is of order $\Delta t$. As before, the coefficients $\veca$, $\vecb$, and $\vecF$ are evaluated at the initial point $(\vecX_t,t)$. Equation~\eqref{eq:dX1} is also called the Milstein method~\cite{book:KloedenPlaten}. The advantage of these stochastic Taylor expansion is that, to order $\Delta t$, one does not need to consider the full continuous-time Brownian trajectory $\{ \vecW_{s}\}_{s=t}^{t+\Delta t}$. Instead, it is sufficient to consider only two random variables, $\Delta \vecW_t$ and $\Delta \vecY_t$, derived from it. Note, however, that they are not independent and $\Delta \vecY_t$ is not Gaussian.

\subsection{Higher-order methods}

Lastly, repeating the procedure above while keeping higher-order terms, one may obtain the order-$\Delta t^{3/2}$ expansion~\cite{book:KloedenPlaten},
\begin{equation}
    \Delta X^\mu_t=a^\mu\Delta t + b^{\mu\nu}\Delta W^\nu_{t}+F^{\mu\nu\sigma}\Delta Y_t^{\nu\sigma}
    +\rho^{\mu\nu} \Delta R_t^\nu + E^{\mu\nu} \Delta L_t^\nu
    +C^{\mu\nu\sigma\rho}\Delta Z_t^{\nu\sigma\rho} + \tau^\mu\Delta t^2+\calo(\Delta t^{2})\,,\label{eq:dX3/2}
\end{equation}
where $\Delta R_t^\nu  = \int_t^{t+\Delta t}\d t'\int_t^{t'}\d W_{t''}^\nu$, $\Delta L_t^\nu = \int_t^{t+\Delta t}\d W^\nu_{t'}\cdot\int_t^{t'}\d t''$, and $\Delta Z_t^{\nu\sigma\rho} = \int_{t}^{t+\Delta t}\d W^\nu_{t'}\cdot\int_{t}^{t'}\d W^\sigma_{t''}\cdot\int_{t}^{t''}\d W^\rho_{t'''}$ are random variables of order $\Delta t^{3/2}$, which too are correlated among themselves and $\Delta\vecW_t$ and $\Delta\vecY_t$, of which $\Delta\vecZ_t$ is also not Gaussian.
In Eq.~\eqref{eq:dX3/2}, the order-$\Delta t^{1/2}$ and -$\Delta t$ expansion coefficients are known from the previous, lower-order expansions [Eqs.~\eqref{eq:dX1/2} and~\eqref{eq:dX1}]. The higher-order coefficients are $\rho^{\mu\nu}=b^{\sigma\nu}\nabla^\sigma a^\mu$, $E^{\mu\nu}=[\partial/\partial t+a^\sigma\nabla^\sigma+D^{\sigma\rho}\nabla^\sigma\nabla^\rho]b^{\mu\nu}$, $C^{\mu\nu\sigma\rho}=b^{\alpha\rho}\nabla^\alpha [b^{\beta\sigma}\nabla^\beta b^{\mu\nu}]$, and $\tau^\mu=(1/2)[\partial/\partial t+a^\sigma\nabla^\sigma+D^{\sigma\rho}\nabla^\sigma\nabla^\rho]a^{\mu}$. All seven coefficients are evaluated, as always, at the initial point $(\vecX_t,t)$. As before, the order-$\Delta t^2$ term is included for proper global convergence~\cite{book:KloedenPlaten}, and hence can be dropped if one is after the Taylor expansion of the propagator rather than consistent approximations of the SDE. We note that surprisingly, as will be shown below, the last four terms happen to cancel out in the particular case of entropy production. 

\subsection{Multiple stochastic integrals}\label{sec:discmult}

Overall, the expressions above show that starting from the order-$\Delta t$ expansion, the displacements $\Delta \vecX_t$ depend on a finite set of correlated random variables, all of which are integrals of the full, continuous-time Wiener process that defines the SDE, $\{\vecW_{s}\}_{s=t}^{t+\Delta t}$. This can be seen directly from the discretized view of the above stochastic integrals:
\begin{equation}
    \Delta W^\nu_{t}=\lim_{\delta t\to0}\sum_{i=1}^{K}\delta W^\nu_i\,,\label{eq:dWt_disc} 
\end{equation}
where $\delta \vecW_i\equiv\vecW_{t+i\delta t}-\vecW_{t+(i-1)\delta t}$ and $K=\Delta t/\delta t$ throughout;
\begin{equation}
    \Delta Y^{\nu\sigma}_t=\lim_{\delta t\to0}\sum_{i=1}^{K}\delta W^\nu_i\sum_{j=1}^{i-1}\delta W^\sigma_j \, ,\label{eq:dYt_disc}
\end{equation}
where we used the It\^o product among differentials; and
\begin{gather}
    \Delta R^\nu_t=\lim_{\delta t\to0}\sum_{i=1}^{K}\delta t\sum_{j=1}^{i-1}\delta W^\nu_j\,,\label{eq:dRt_disc}\\
    \Delta L^\nu_t=\lim_{\delta t\to0}\sum_{i=1}^{K}\delta W^\nu_i\sum_{j=1}^{i-1}\delta t\,,\label{eq:dLt_disc}\\
    \Delta Z^{\nu\sigma\rho}_t=\lim_{\delta t\to0}\sum_{i=1}^{K}\delta W^\nu_i\sum_{j=1}^{i-1}\delta W^\sigma_j\sum_{k=1}^{j-1}\delta W^\rho_k \, .\label{eq:dZt_disc}
\end{gather}
Note that the means of all these five integrals (sums) are zero~\cite{book:KloedenPlaten}, since the inner sums over time stop before they reach the the Wiener increment of the outer sum's time increment. Hence, they are independent and $\langle\delta\vecW_i\rangle=\veco$.
The aim of these multiple stochastic integrals is to convert $K\to\infty$ integrals over $\lim_{\delta t\to0}\{\delta \vecW_i\}_{i=1}^{K}$ to a few integrals over these parameters, whose moments could be computes from the discretized expressions. 

\section{Consistent Propagator expansions}\label{sec:propagator}

In this Section, we present the approach we propose to consistently find high-order propagators. It utilizes the setup and notation of Sec.~\ref{sec:model}, and the method of stochastic Taylor expansions of Sec.~\ref{sec:expans}.

\subsection{The leading-order propagator}\label{sec:propSIMP}

Prior to going into the detailed derivation of the propagator expansion to any order, we recall the simplest propagator. It is obtained by a na\"ive replacement of $\d t\to\Delta t$ in Eq.~\eqref{eq:SDE},
\begin{equation}
    \Delta\vecX_t\simeq\veca(\vecX_t,t)\Delta t+\vecb(\vecX_t,t)\cdot\Delta\vecW_t \,.\label{eq:dXnaive}
\end{equation}
Here, we replaced $\d\vecW_t\to\Delta\vecW_t$ as well (a Wiener increment of variance $\vecI\Delta t$, which is independent of $\vecX_t$). As we saw in Sec.~\ref{sec:expans}, it corresponds to the lowest-order approximation of the SDE called the Euler-Maruyama scheme [Eq.~\eqref{eq:dX1/2}]. 

With this replacement, $\Delta\vecX_t$ only depends on $\Delta\vecW_t$ (as opposed to the Milstein and higher-order schemes). Thus, we can invert the relation between them, and express the noise magnitude $\Delta\vecW_t$ that should have acted at point $(\vecX_t,t)$ to obtain the displacement $\Delta\vecX_t$, solely in terms of the latter,
\begin{equation}
    \Delta\vecW_t\simeq\vecb^{-1}(\vecX_t,t)\cdot[\Delta\vecX_t-\veca(\vecX_t,t)\Delta t]\,.\label{eq:dWtdXt_inv}
\end{equation}
Equation~\eqref{eq:dWtdXt_inv} is no more than just a change of variables from a random variable $\Delta\vecX_t$ to a random variable $\Delta\vecW_t$.

This inversion is useful since we may now utilize the exactly-normal distribution of $\Delta\vecW_t$. Since only $\Delta\vecW_t$ dictates the displacement $\Delta\vecX_t$ given the initial state $(\vecX_t,t)$ in the Euler-Maruyama approximation, via probability conservation, we write $P(\vecx+\Delta\vecx,t+\Delta t|\vecx,t)\d\Delta\vecx=\Pr(\Delta\vecW_t=\Delta\vecw)\d\Delta\vecw$, where $\Pr(\Delta\vecW_t=\Delta\vecw)=(2\pi\Delta t)^{-N/2}\exp[-|\Delta\vecw|^2/(2\Delta t)]$ and $\Delta\vecw=\vecb^{-1}(\vecx,t)\cdot[\Delta\vecx-\veca(\vecx,t)\Delta t]$. With the Jacobian $\d\Delta\vecw/\d\Delta\vecx=\vecb^{-1}(\vecx,t)$, we find the order-$\Delta t^{1/2}$ propagator, which is the probability distribution function describing the Euler-Maruyama scheme below~\cite{book:schuss},
\begin{equation}
    P_{1/2}(\vecx+\Delta\vecx,t+\Delta t|\vecx,t)=\frac{1}{\sqrt{\det[4\pi\vecD(\vecx,t)\Delta t]}}\exp\left\{-\frac{\vecD^{-1}(\vecx,t)}{4\Delta t}:[\Delta\vecx-\veca(\vecx,t)\Delta t][\Delta\vecx-\veca(\vecx,t)\Delta t]\right\}\,.\label{eq:propSIMP}
\end{equation}
By ``order-$\Delta t^{1/2}$ propagator'', we mean that the dynamics generated by Eq.~\eqref{eq:propSIMP} agree with trajectories of the exact SDE Eq.~\eqref{eq:langevin} to this order.

\subsection{Computing the exact propagator}

If the complete expression for $p(\vecx+\Delta\vecx,t+\Delta t|\vecx,t)$ is known, one can perform a Taylor expansions in small $\Delta\vecx \sim \Delta t^{1/2}$ (as we do in Appendix~\ref{app:GBM}). The leading order has to be Eq.~\eqref{eq:propSIMP}. As we explained in Sec.~\ref{sec:discmult}, for any scheme beyond leading order, $\Delta\vecX_t$ is a function of more than one random variable, and these variables are also correlated. Thus, we follow an alternative route to Sec.~\ref{sec:propSIMP}.

For any arbitrarily (long or short) time duration $\Delta t$, the propagator, by definition, is given by
\begin{equation}
    P(\vecx+\Delta\vecx,t+\Delta t|\vecx,t)=\left\langle\delta\left[\Delta\vecx-\int_t^{t+\Delta t}\veca(\vecX_{t'},t')\d t'-\int_t^{t+\Delta t}\vecb(\vecX_{t'},t')\cdot\d\vecW_{t'}\right]\Bigg|\vecX_t=\vecx\right\rangle \,.\label{eq:propDELTA}
\end{equation}
The delta function imposes the stochastic equation of motion [Eq.~\eqref{eq:langevin}] upon the realization of the random displacement $\Delta\vecX_t=\Delta\vecx$, given the initial condition $\vecX_t=\vecx$. We Fourier-transform to get its characteristic function, $\tilde{P}(\vecq,\Delta t|\vecx,t)=\int\d\Delta\vecx e^{i\vecq\cdot\Delta\vecx}P(\vecx+\Delta\vecx,t+\Delta t|\vecx,t)$, which becomes
\begin{equation}
    \tilde{P}(\vecq,\Delta t|\vecx,t)=\left\langle\exp\left\{i\vecq\cdot\left[\int_t^{t+\Delta t}\veca(\vecX_{t'},t')\d t'+\int_t^{t+\Delta t}\vecb(\vecX_{t'},t')\cdot\d\vecW_{t'}\right]\right\}\Bigg|\vecX_t=\vecx\right\rangle \,.\label{eq:propEXP}
\end{equation}
The ensemble average in both Eqs.~\eqref{eq:propDELTA} and~\eqref{eq:propEXP} is over all noise realizations, $\{\delta \vecW_s\}_{s=t}^{t+\Delta t}$ (that is, we do not average over the initial point $\vecX_t=\vecx$). So far, both equations are exact.

Another exact construction we may introduce is the Wiener discrete path integral~\cite{book:schuss}, namely, the probability distribution to obtain a collection of discrete-time noise realizations, $\{\delta \vecW_i\}_{i=1}^K$. It is given by the following product of $K$ Gaussians,
\begin{equation}
    \Pr\left[\{\delta \vecW_i=\delta\vecw_i\}_{i=1}^{K}\right]=\prod_{i=1}^K\left[\frac{1}{(2\pi\delta t)^{N/2}}\exp\left(-\frac{|\delta\vecw_i|^2}{2\delta t}\right)\right]\,.\label{eq:MBM}
\end{equation}
Note that the increments of the Wiener process are normally distributed, so this decomposition is exact regardless of how small $\delta t$ is. Therefore, while we aim to reconstruct the path integral for the positions $\vecX_t$ while paying close attention to whether we kept sufficient high-order correction, Eq.~\eqref{eq:MBM} is formal and always exact~\cite{book:schuss}. The ensemble averages appearing in Eqs.~\eqref{eq:propDELTA} and~\eqref{eq:propEXP} are defined by the integration using the probability measure of Eq.~\eqref{eq:MBM}.

We define the following functionals of the Wiener process, conditioned on the initial conditions,
\begin{eqnarray}
    {\Delta\boldsymbol{\mathcal{D}}}_t &=& \Delta\vecX_t\left[\{\vecW_{t'}\}_{t'=t}^{t+\Delta t}|\vecX_t,t\right] -  \veca (\vecX_t,t) \Delta t - \vecb (\vecX_t,t) \cdot\Delta \vecW_{t} \, ,\nonumber \\
     \error_t &=& {\Delta\boldsymbol{\mathcal{D}}}_t - \langle {\Delta\boldsymbol{\mathcal{D}}}_t |\vecX_t,t \rangle \,.\label{eq:error0}
\end{eqnarray}
In Eq.~\eqref{eq:error0}, ${\Delta\boldsymbol{\mathcal{D}}}_t$ measures the difference between the exact $\Delta\vecX_t$ [Eq.~\eqref{eq:langevin}] and the Euler-Maruyama approximation [Eq.~\eqref{eq:dX1/2}], and $\error_t$ is its centered difference. Explicitly, $\error_t$ is given by
\begin{equation}
    \error_t\left[\{\vecW_{t'}\}_{t'=t}^{t+\Delta t}|\vecX_t,t\right]=\int_t^{t+\Delta t}\veca(\vecX_{t'},t')\d t'+\int_t^{t+\Delta t}\vecb(\vecX_{t'},t')\cdot\d\vecW_{t'}-\langle\Delta\vecX_t|\vecX_t,t\rangle-\vecb(\vecX_t,t)\cdot\Delta\vecW_t \,.\label{eq:error}
\end{equation}
The notation stresses that $\error_t$ is defined on the probability space of the increments of the Wiener process $\{\vecW_{t'}\}_{t'=t}^{t+\Delta t}$ and is conditioned on $\vecX_t$.

For small $\Delta t$, according to Eq.~\eqref{eq:dX3/2}, $\langle\Delta\vecX_t|\vecX_t\rangle=\veca(\vecX_t,t)\Delta t+\boldsymbol\tau(\vecX_t,t)\Delta t^2+\calo(\Delta t^3)$, and $\neverror^\mu_t=F^{\mu\nu\sigma}\Delta Y_t^{\nu\sigma}+\rho^{\mu\nu}\Delta R_t^\nu + \lambda^{\mu\nu}\Delta L_t^\nu+C^{\mu\nu\sigma\rho}\Delta Z_t^{\nu\sigma\rho} +\calo(\Delta t^2)$. Thus, $\error_t=\calo(\Delta t)$ can be brought to depend explicitly on $\{\vecW_{s}\}_{s=t}^{t+\Delta t}$ via the multiple stochastic integrals (arising from the expansion schemes of Sec.~\ref{sec:expans} and beyond~\cite{book:KloedenPlaten}) and on the initial position $\vecX_t$.


Combining Eqs.~\eqref{eq:propEXP}, \eqref{eq:MBM}, \eqref{eq:error}, and~\eqref{eq:dWt_disc}, we arrive at the $\delta t$-discretized path integral,
\begin{eqnarray}
    \tilde{P}(\vecq,\Delta t|\vecx,t)&=&\lim_{\delta t\to0}\prod_{i=1}^K\left[\int\frac{\d\delta\vecw_i}{(2\pi\delta t)^{N/2}}\right]\nonumber\\
    &\times&\exp\left\{-\sum_{j=1}^K\frac{|\delta\vecw_j|^2}{2\delta t}+i\vecq\cdot\left[\error_t[\{\delta \vecw_j\}|\vecx,t]+\langle\Delta\vecX_t|\vecx,t\rangle+\vecb(\vecx,t)\cdot\sum_{j=1}^K\delta\vecw_j\right]\right\}\,.
\end{eqnarray}
Completing the square,
\begin{equation}
    \frac{|\delta\vecw_j|^2}{2\delta t}-i\vecq\cdot\vecb\cdot\delta\vecw_j=\frac{|\delta\vecw_j-i\vecq\cdot\vecb\delta t|^2}{2\delta t}+\vecq\cdot\vecD\cdot\vecq\delta t \,,
\end{equation}
with $\sum_{i=1}^K\delta t=\Delta t$, we get the expression,
\begin{eqnarray}
    \tilde{P}(\vecq,\Delta t|\vecx,t)&=&e^{-\vecq\cdot\vecD(\vecx,t)\cdot\vecq\Delta t+i\vecq\cdot\langle\Delta\vecX_t|\vecx,t\rangle}\nonumber\\
    &\times&\lim_{\delta t\to0}\prod_{i=1}^K\left[\int\frac{\d\delta\vecw_i}{(2\pi\delta t)^{N/2}}\exp\left(-\frac{|\delta\vecw_i-i\vecq\cdot\vecb(\vecx,t)\delta t|^2}{2\delta t}\right)\right]e^{i\vecq\cdot\error_t[\{\delta \vecw_i\}|\vecx,t]} \,. \label{eq:PtildeGauss}
\end{eqnarray}
This is the average of  $e^{i\vecq\cdot\error_t}$ for a Langevin process whose Wiener increments have nonzero means.
Changing variables $\delta\tilde\vecw_j=\delta\vecw_j-i\vecq\cdot\vecb\delta t$ to return to the usual (zero-mean) Wiener process, we find the basis for the incoming expansions,
\begin{equation}
    \tilde{P}(\vecq,\Delta t|\vecx,t)=e^{-\vecq\cdot\vecD(\vecx,t)\cdot\vecq\Delta t+i\vecq\cdot\langle\Delta\vecX_t|\vecx,t\rangle}\langle e^{i\vecq\cdot\error_t[\{\vecW_{t'}+i\vecq\cdot\vecb(\vecx,t)t'\}|\vecx,t]}\rangle\,,\label{eq:propGEN}
\end{equation}
where the expectation is once again over the standard measure of Wiener process, so $\error$ has absorbed the offset $i\vecq\cdot\vecb\delta t$. Note that $\vecb$ is evaluated at time $t$ and not $t'$ and hence the offset is a constant within the SDE, because this term comes from the Euler-Maruyama approximation.
This is the first central result of this paper, and is completely exact so far. In fact, there is no discretization in Eq.~\eqref{eq:propGEN}, as $\error$ could be expressed with integrals over time (see Eq.~\eqref{eq:error}).

We now proceed to consider small $\Delta t$, which allows systematic expansions. From the Gaussian of Eq.~\eqref{eq:propGEN}, we see that $\vecq\sim\Delta t^{-1/2}$. Combined with $\error_t\sim\Delta t$ [see discussion below Eq.~\eqref{eq:error}], we see that $\vecq\cdot\error_t\sim\Delta t^{1/2}$ is a small quantity, so we may (stochastic) Taylor-expand the exponential within the ensemble average in Eq.~\eqref{eq:propGEN} for small $\vecq\cdot\error_t$.

\subsection{Euler-Maruyama propagator}\label{sec:prop1/2}

We begin by obtaining the leading-order propagator of the Euler-Maruyama (order-$\Delta t^{1/2}$) method, Eq.~\eqref{eq:dX1/2}. By definition [Eq.~\eqref{eq:error}], we have that
\begin{eqnarray}
    \langle\Delta\vecX_t|\vecx,t\rangle&=&\veca(\vecx,t)\Delta t+\calo(\Delta t^2)\,,\nonumber\\
    \error_t[\{\vecW_{t'}\}|\vecx,t]&=&\calo(\Delta t)\,.
\end{eqnarray}
Thus, $\langle e^{-i\vecq\cdot\error_t[\{\vecW_{t'}+i\vecq\cdot\vecb t'\}|\vecx,t]}\rangle=1+\calo(\Delta t^{1/2})$. To leading order we are left with,
\begin{equation}
    \tilde{P}(\vecq,\Delta t|\vecx,t)=e^{-\vecq\cdot\vecD\cdot\vecq\Delta t+i\vecq\cdot\veca\Delta t}[1+\calo(\Delta t^{1/2})]\,,
\end{equation}
which is a Fourier transform of a Gaussian. Upon inverting it, $P(\vecx+\Delta\vecx,t+\Delta t|\vecx,t)=(2\pi)^{-N}\int\d\vecq e^{-i\vecq\cdot\Delta\vecx}\tilde{P}(\vecq,\Delta t|\vecx,t)$, we reobtain Eq.~\eqref{eq:propSIMP},
\begin{equation}
    P(\vecx+\Delta\vecx,t+\Delta t|\vecx,t)=P_{1/2}(\vecx+\Delta\vecx,t+\Delta t|\vecx,t)[1+\calo(\Delta t^{1/2})]\,.\label{eq:prop1/2}
\end{equation}
However, now we explicitly know the discrepancy among the approximate propagator of Eq.~\eqref{eq:dX1/2} and the exact propagator of Eq.~\eqref{eq:langevin}. 

As briefly discussed in Sec.~\ref{sec:EulerMaruyamanMethod}, most of the literature on stochastic Taylor expansions~\cite{book:KloedenPlaten} aims to obtain numerical integration methods of SDEs and their accuracy, rather than express the errors of the propagator expansion as we do in Eq.~\eqref{eq:prop1/2}. This is done for a good reason\,---\,expanding the propagator itself obscures the so-called strong approximation error~\cite{book:KloedenPlaten}. The order of strong convergence is the `true' error resulting from using the approximate Euler-Maruyama method (Eq.~\eqref{eq:dX1/2}, and its corresponding propagator, Eq.~\eqref{eq:prop1/2}) to sample trajectories from the SDE, instead of the exact (unknown) propagator $P$. 
This is best understood as follows. Consider $\vecb=\veco$, where the Euler-Maruyama propagator is the same as the leading-order Euler approximation for the ordinary differential equation $\dot\vecX(t)=\veca(\vecX(t),t)$. The Euler approximation is correct to order $\Delta t^2$ (called the local truncation error~\cite{book:Iserles}). Taking $\sim\Delta t^{-1}$ steps builds up to a global error of order $\Delta t$, which is why the Euler method for ordinary differential equations is referred to as order $\Delta t$.\footnote{A more rigorous analysis, taking into account the buildup of errors across subsequent steps, also leads to an $\mathcal{O} (\Delta t)$ global error estimate, but with a prefactor that grows exponentially in the integration time~\cite{book:Iserles}.} Going back to the Euler-Maruyama method, upon subtracting the sufficiently accurate drift, the error in the trajectory is $\error_t$, which has zero mean and variance $\Delta t^2$. Hence, the error in the variance is $\Delta t^2$. Due to the Markov property, the errors of subsequent steps in the Langevin process are independent, building up to a global error which once again has variance of order $\Delta t$. The error in the trajectory is proportional to the standard deviation, which is therefore $\calo(\Delta t^{1/2})$. This is why the Euler-Maruyama approximation is referred to a numerical scheme of (strong) order $\Delta t^{1/2}$. 

In this paper, we are interested in obtaining expansions of the propagator to prescribed orders in $\Delta t$, to which end the notation in Eq.~\eqref{eq:prop1/2} is convenient as it separates the leading-order propagator $P_{1/2}$ and the next order terms, which are smaller by factors of order $\Delta t^{1/2}$. This difference between our efforts and the mathematical literature manifests in the expression for the entropy production, which requires substituting trajectories $\vecx\to\vecx+\Delta\vecx$ (that were already sampled elsewhere) into the (unknown) explicit functional of the propagator. The accuracy of the expression for the propagator itself, albeit important to stochastic thermodynamics, is a new issue in the context of solving stochastic differential equations, and thus is not addressed in the literature on numerical schemes~\cite{book:KloedenPlaten}.

\subsection{Milstein propagator}

We proceed to evaluate the first nontrivial propagator, representing the Milstein (order-$\Delta t$) method, Eq.~\eqref{eq:dX1}. Now we have
\begin{eqnarray}
    \langle\Delta\vecX_t|\vecx,t\rangle&=&\veca(\vecx,t)\Delta t+\calo(\Delta t^2)\,,\nonumber\\
    \neverror_t^\mu[\{\vecW_{t'}\}|\vecx,t]&=&F^{\mu\nu\sigma}(\vecx,t)\Delta Y^{\nu\sigma}_t[\{\vecW_{t'}\}]+\calo(\Delta t^{3/2})\, . \label{eq:error_order1}
\end{eqnarray}
We Taylor expand in small $\vecq\cdot\error_t$ to include the first order term,
\begin{equation}
    e^{i\vecq\cdot\error_t[\{\vecW_{t'}+i\vecq\cdot\vecb t'\}|\vecx,t]}=1+i\vecq\cdot\error_t[\{\vecW_{t'}+i\vecq\cdot\vecb t'\}|\vecx,t]+\calo(\Delta t)\,.
\end{equation}
Thus, we should compute $\error_t$ to the required accuracy \eqref{eq:error_order1}, with a Gaussian process whose increments are offset, $\d\vecW_{t'}\to\d\vecW_{t'}+i\vecq\cdot\vecb(\vecx,t)\d t'$. The expression for $\Delta Y$ for the offset process is obtained through Eq.~\eqref{eq:dYt_disc},
\begin{eqnarray}
    \Delta Y^{\nu\sigma}_t[\{\vecW_{t'}+i\vecq\cdot\vecb t'\}]&=&
    \int_t^{t+\Delta t}\d W^\nu_{t'} \int_t^{t'}\d W^\sigma_{t''}+(iq^\alpha) b^{\alpha\nu}\int_t^{t+\Delta t}\d t' \int_t^{t'}\d W^\sigma_{t''}\nonumber\\
    &+&(iq^\alpha) b^{\alpha\sigma}\int_t^{t+\Delta t}\d W^\nu_{t'} \int_t^{t'}\d t''+\frac 12(iq^\alpha )(iq^\beta)b^{\alpha\nu}b^{\beta\sigma}\Delta t^2\,.\label{eq:dYt_off}
\end{eqnarray}
According to Eq.~\eqref{eq:dYt_off} [remembering the zero mean of Eqs.~\eqref{eq:dYt_disc}, \eqref{eq:dRt_disc}, and~\eqref{eq:dLt_disc}], we have,
\begin{equation}
    \langle \Delta Y^{\nu\sigma}_t[\{\vecW_{t'}+i\vecq\cdot\vecb t'\}]\rangle=\frac 12(iq^\alpha )(iq^\beta)b^{\alpha\nu}b^{\beta\sigma}\Delta t^2\,.
\end{equation}
Recalling that $F^{\mu\nu\sigma}=b^{\rho\sigma}\nabla^\rho b^{\mu\nu}$ (see Eq.~\eqref{eq:dX1}), we find,
\begin{equation}
    \tilde{P}(\vecq,\Delta t|\vecx,t)=e^{-\vecq\cdot\vecD\cdot\vecq\Delta t+i\vecq\cdot\veca\Delta t}[1+(iq^\mu )(iq^\alpha )(iq^\beta)b^{\alpha\nu}D^{\rho\beta}\nabla^\rho b^{\mu\nu}\Delta t^2+\calo(\Delta t)]\,.
\end{equation}
We get the leading order Gaussian, and also a cubic term in $\vecq$. Note that after summing over $\mu$ and $\alpha$, only the symmetric part in $\mu\leftrightarrow\alpha$ survives within the coefficient, $q^\mu q^\alpha b^{\alpha\nu}\nabla b^{\mu\nu}=(1/2)q^\mu q^\alpha [b^{\alpha\nu}\nabla b^{\mu\nu}+b^{\mu\nu}\nabla b^{\alpha\nu}]=q^\mu q^\alpha \nabla D^{\mu\alpha}$. With that, using the inverse Fourier transform ($i\vecq\to(-\partial/\partial\Delta\vecx)$), we get
\begin{equation}
    P(\vecx+\Delta\vecx,t+\Delta t|\vecx,t)=\left[1-D^{\rho\sigma}\left(\nabla^\rho D^{\mu\nu}\right)\Delta t^2\frac{\partial}{\partial\Delta x^\mu}\frac{\partial}{\partial\Delta x^\nu}\frac{\partial}{\partial\Delta x^\sigma}+\calo(\Delta t)\right]P_{1/2}(\vecx+\Delta\vecx,t+\Delta t|\vecx,t)\,.\label{eq:prop1der}
\end{equation}

We proceed to compute Eq.~\eqref{eq:prop1der} explicitly by taking the three derivatives of the Gaussian Eq.~\eqref{eq:propSIMP}, and keeping only order-$\Delta t^{1/2}$ terms. Since this involves several derivatives of a Gaussian, it is convenient to work with the multidimensional generalization of the Hermite polynomials. Losely speaking, this is the reason why a direct expansion of the Fokker-Planck propagator is the most useful in this basis~\cite{AitSahaliaECON02,AitSahaliaAT08}. Based upon the Rodrigues formula~\cite{BOOK:szego1975}, we define
\begin{equation}
    \vecH_n(\vecy;\vecM)=(-1)^ne^{\vecy\cdot\vecM^{-1}\cdot\vecy/2}\frac{\partial^n}{\partial\vecy^n}e^{-\vecy\cdot\vecM^{-1}\cdot\vecy/2}=\left(\vecM^{-1}\cdot\vecy-\frac{\partial}{\partial\vecy}\right)^n\cdot1\label{eq:Hermite}
\end{equation}
for some vector $\vecy$ and a constant scaling tensor $\vecM$. The $n$th hermite polynomial here is a rank-$n$ tensor. In terms of $\vecH_n$, the propagator reads,
\begin{equation}
    P(\vecx+\Delta\vecx,t+\Delta t|\vecx,t)=P_{1/2}(\vecx+\Delta\vecx,t+\Delta t|\vecx,t)\left[1+D^{\rho\sigma}\left(\nabla^\rho D^{\mu\nu}\right) H_3^{\mu\nu\sigma}(\Delta\vecx,2\vecD\Delta t)\Delta t^2+\calo(\Delta t)\right]\,.\label{eq:prop1herm}
\end{equation}
We note that $\vecH_n(\Delta\vecx,2\vecD\Delta t)\sim\Delta t^{-n/2}$. Explicitly, the propagator reads,
\begin{eqnarray}
    P(\vecx+\Delta\vecx,t+\Delta t|\vecx,t)&=&P_{1/2}(\vecx+\Delta\vecx,t+\Delta t|\vecx,t)\nonumber\\
    &\times&\left\{1+\frac18[\nabla^\sigma (\vecD^{-1})^{\mu\nu}]\left( 2D^{\mu\nu}\Delta x^{\sigma}+ 4D^{\nu\sigma}\Delta x^{\mu}-\frac{\Delta x^{\mu}\Delta x^{\nu}\Delta x^{\sigma}}{\Delta t}\right)+\calo(\Delta t)\right\}\,,\label{eq:prop1}
\end{eqnarray}
where we used the identity $\nabla(\vecD^{-1})=-\vecD^{-1}\cdot\nabla\vecD\cdot\vecD^{-1}$ and the symmetry of $\vecD$ to replace $\grad(\vecD^{-1})^{\mu\nu}(D^{\mu\sigma}\Delta x^\nu+D^{\nu\sigma}\Delta x^\mu)=2\grad(\vecD^{-1})^{\mu\nu}D^{\nu\sigma}\Delta x^\mu$. In Eqs.~\eqref{eq:prop1der}, Eq.~\eqref{eq:prop1herm}, and~\eqref{eq:prop1}, all $\vecD=\vecb\cdot\vecb^\T/2$ are evaluated at $(\vecx,t)$, while the dependence on $\Delta\vecx$ appears explicitly. 

Equation~\eqref{eq:prop1} is the first, beyond-Gaussian propagator that one can find via the scheme we propose. Since all expansions were performed consistently, there is no dependence on discretization. In the notation of Eq.~\eqref{eq:propSTRUC}, we identify
\begin{equation}
    \Phi^\rho(\vecK|\vecx,t)=\frac14[\nabla^\rho (\vecD^{-1})^{\mu\nu}(\vecx,t)]D^{\mu\nu}(\vecx,t)+\frac12 [\nabla^\sigma(\vecD^{-1})^{\rho\nu}(\vecx,t)]D^{\nu\sigma}(\vecx,t)-\frac1{8}[\nabla^\rho  (\vecD^{-1})^{\mu\nu}(\vecx,t)]K^{\mu\nu}\,,\label{eq:phi}
\end{equation}
where $\vecK=\Delta\vecx\Delta\vecx/\Delta t$. We remark that to order $\Delta t^{1/2}$, $\Psi(\vecK,\vecx,t)=0$ in Eq.~\eqref{eq:propSTRUC}.

\subsection{Higher-order propagator}

We finish by calculating the propagator corresponding to the order-$\Delta t^{3/2}$ method, Eq.~\eqref{eq:dX3/2}. It requires a higher power of $\error_t$, which involves a more elaborate calculation. First, combining Eqs.~\eqref{eq:error} and~\eqref{eq:dX3/2}, we identify
\begin{eqnarray}
    \langle\Delta\vecX_t|\vecx,t\rangle&=&\veca(\vecx,t)\Delta t+\boldsymbol\tau(\vecx,t)\Delta t^2+\calo(\Delta t^3)\,,\nonumber\\
    \neverror_t^\mu[\{\vecW_{t'}\}|\vecx,t]&=&F^{\mu\nu\sigma}(\vecx,t)\Delta Y^{\nu\sigma}_t[\{\vecW_{t'}\}]+\rho^{\mu\nu} (\vecx,t)\Delta R_t^\nu[\{\vecW_{t'}\}] + E^{\mu\nu} (\vecx,t)\Delta L_t^\nu[\{\vecW_{t'}\}]
    \nonumber\\&+&C^{\mu\nu\sigma\rho}(\vecx,t)\Delta Z_t^{\nu\sigma\rho}[\{\vecW_{t'}\}]+\calo(\Delta t^2)\,.
\end{eqnarray}
Next, 
\begin{equation}
    e^{i\vecq\cdot\error_t[\{\vecW_{t'}+i\vecq\cdot\vecb t'\}|\vecx,t]}=1+i\vecq\cdot\error_t[\{\vecW_{t'}+i\vecq\cdot\vecb t'\}|\vecx,t]+\frac12\{i\vecq\cdot\error_t[\{\vecW_{t'}+i\vecq\cdot\vecb t'\}|\vecx,t]\}^2+\calo(\Delta t^{3/2})\,, 
\end{equation}
where only $\Delta\vecY\Delta\vecY$ in the quadratic term contributes to the required order. Offseting the increments on the Wiener process by $\d\vecW_{t'}\to\d\vecW_{t'}+i\vecq\cdot\vecb(\vecx,t)\d t'$, in Eqs.~\eqref{eq:dRt_disc}, \eqref{eq:dLt_disc}, and~\eqref{eq:dZt_disc}, we find the ensemble averages
\begin{eqnarray}
    \langle\Delta R^{\nu}_t[\{\vecW_{t'}+i\vecq\cdot\vecb t'\}]\rangle&=&
    \frac12(iq^\alpha) b^{\alpha\nu}\Delta t^2\,,\nonumber\\
    \langle\Delta L^{\nu}_t[\{\vecW_{t'}+i\vecq\cdot\vecb t'\}]\rangle&=&
    \frac12(iq^\alpha) b^{\alpha\nu}\Delta t^2\,,\nonumber\\
    \langle\Delta Z^{\nu\sigma\rho}_t[\{\vecW_{t'}+i\vecq\cdot\vecb t'\}]\rangle&=&
    \frac 16(iq^\alpha )(iq^\beta)(iq^\gamma)b^{\alpha\nu}b^{\beta\sigma}b^{\gamma\rho}\Delta t^3\,.
\end{eqnarray}
From Eq.~\eqref{eq:dYt_off},
\begin{eqnarray}
    & &\langle\Delta Y^{\nu\sigma}_t[\{\vecW_{t'}+i\vecq\cdot\vecb t'\}]\Delta Y^{\nu'\sigma'}_t[\{\vecW_{t'}+i\vecq\cdot\vecb t'\}]\rangle=\frac12\delta^{\nu\nu'}\delta^{\sigma\sigma'}\Delta t^2\nonumber\\
    & &+\left[\frac13(iq^\alpha)(iq^{\alpha'}) b^{\alpha\nu}b^{\alpha'\nu'}\delta^{\sigma\sigma'}+\frac13(iq^\beta)(iq^{\beta'}) b^{\beta\sigma}b^{\beta'\sigma'}\delta^{\nu\nu'}+\frac16(iq^\alpha)(iq^{\beta'}) b^{\alpha\nu}b^{\beta'\sigma'}\delta^{\sigma\nu'}+\frac16(iq^\beta)(iq^{\alpha'}) b^{\beta\sigma}b^{\alpha'\nu'}\delta^{\nu\sigma'}\right]\Delta t^3\nonumber\\
    &&+\frac 14(iq^\alpha )(iq^{\alpha'} )(iq^\beta)(iq^{\beta'})b^{\alpha\nu}b^{\alpha'\nu'}b^{\beta\sigma}b^{\beta'\sigma'}\Delta t^4,
\end{eqnarray}
where only terms with even powers of $\vecW_{t'}$ survived the average.

It total, keeping only symmetric terms for repeating powers of $\vecq$, and taking the inverse Fourier-transform, we find
\begin{eqnarray}
    P(\vecx+\Delta\vecx,t+\Delta t|\vecx,t)&=&\left[1-D^{\rho\sigma}\nabla^\rho D^{\mu\nu}\Delta t^2\frac{\partial}{\partial\Delta x^\mu}\frac{\partial}{\partial\Delta x^\nu}\frac{\partial}{\partial\Delta x^\sigma}\right.\nonumber\\
    &+&\psi_{2}^{\mu\nu}\Delta t^2\frac{\partial}{\partial\Delta x^\mu}\frac{\partial}{\partial\Delta x^\nu}+\psi_{4}^{\mu\nu\sigma\rho}\Delta t^3\frac{\partial}{\partial\Delta x^\mu}\frac{\partial}{\partial\Delta x^\nu}\frac{\partial}{\partial\Delta x^\sigma}\frac{\partial}{\partial\Delta x^\rho}\nonumber\\
    &+&\left.\psi_{6}^{\mu\nu\sigma\rho\alpha\beta}\Delta t^4\frac{\partial}{\partial\Delta x^\mu}\frac{\partial}{\partial\Delta x^\nu}\frac{\partial}{\partial\Delta x^\sigma}\frac{\partial}{\partial\Delta x^\rho}\frac{\partial}{\partial\Delta x^\alpha}\frac{\partial}{\partial\Delta x^\beta}+\calo(\Delta t^{3/2})\right]\times\nonumber\\
    &\times&\frac{1}{\sqrt{\det[4\pi\vecD\Delta t]}}\exp\left\{-\frac{\vecD^{-1}}{4\Delta t}:[\Delta\vecx-\veca\Delta t-\boldsymbol\tau\Delta t^2][\Delta\vecx-\veca\Delta t-\boldsymbol\tau\Delta t^2]\right\}\,,\label{eq:prop3/2der}
\end{eqnarray}
where
\begin{eqnarray}
    \psi_{2}^{\mu\nu}&=&D^{\nu\sigma}\nabla^{\sigma}a^{\mu}+\frac{1}{2}\frac{\partial D^{\mu\nu}}{\partial t}+\frac{1}{2}a^{\sigma}\nabla^{\sigma}D^{\mu\nu}+\frac{1}{2}D^{\sigma\rho}\nabla^{\sigma}\nabla^{\rho}D^{\mu\nu} \,,\label{eq:psi2}\\
    \psi_{4}^{\mu\nu\sigma\rho}&=&\frac23D^{\alpha\rho}\nabla^{\alpha}(D^{\beta\sigma}\nabla^{\beta}D^{\mu\nu})+\frac13D^{\alpha\beta}(\nabla^{\beta}D^{\mu\sigma})(\nabla^{\alpha}D^{\nu\rho})\,,\label{eq:psi4}\\
    \psi_{6}^{\mu\nu\sigma\rho\alpha\beta}&=&\frac{1}{2}D^{\beta\gamma}(\nabla^{\gamma}D^{\nu\rho})D^{\alpha\kappa}(\nabla^{\kappa}D^{\mu\sigma})\,.\label{eq:psi6}
\end{eqnarray}
Note that the Gaussian part of Eq.~\eqref{eq:prop3/2der} includes the order-$\Delta t^2$ ``overshoot'' contribution to the mean which does not appear in Eq.~\eqref{eq:propSIMP}. It allows for the mean drift to be accurately computed to order-$\Delta t^2$, that is, to get exactly $\langle\Delta\vecX\rangle=\veca\Delta t+\boldsymbol\tau\Delta t^2+\calo(\Delta t^3)$; see Eq.~\eqref{eq:dX3/2} and the discussion below Eq.~\eqref{eq:prop1/2} on strong convergence. 

The next step again involves expressing the derivatives using the multidimensional Hermite tensors, Eq.~\eqref{eq:Hermite}. 
In Eq.~\eqref{eq:prop1herm}, we dropped $\veca\Delta t$ from the Hermite polynomial $\vecH_3$.
Here we must keep it now in order to get the correct order of approximation. 
The propagator now reads,
\begin{eqnarray}
    P(\vecx+\Delta\vecx,t+\Delta t|\vecx,t)&=&\frac{1}{\sqrt{\det[4\pi\vecD\Delta t]}}\exp\left\{-\frac{\vecD^{-1}}{4\Delta t}:[\Delta\vecx-\veca\Delta t-\boldsymbol\tau\Delta t^2][\Delta\vecx-\veca\Delta t-\boldsymbol\tau\Delta t^2]\right\}\nonumber\\
    &\times&\Big[1+D^{\rho\sigma}\nabla^\rho D^{\mu\nu}H_3^{\mu\nu\sigma}(\Delta\vecx-\veca\Delta t,2\vecD\Delta t)\Delta t^2+\psi_{2}^{\mu\nu}H_2^{\mu\nu}(\Delta\vecx,2\vecD\Delta t)\Delta t^2\nonumber\\
    &+&\psi_{4}^{\mu\nu\sigma\rho}H_4^{\mu\nu\sigma\rho}(\Delta\vecx,2\vecD\Delta t)\Delta t^3+\psi_{6}^{\mu\nu\sigma\rho\alpha\beta}H_6^{\mu\nu\sigma\rho\alpha\beta}(\Delta\vecx,2\vecD\Delta t)\Delta t^4+\calo(\Delta t^{3/2})\Big] \,.\label{eq:prop3/2herm}
\end{eqnarray}
The additional offset $\veca\Delta t$ in $\vecH_3$ translates into another order-$\Delta t$ term of the form $\vecH_2\Delta t^2(\Delta\vecx,2\vecD\Delta t)$, while the order-$\Delta t^{1/2}$ term is the same as in Eq.~\eqref{eq:prop1herm},
\begin{eqnarray}
    P(\vecx+\Delta\vecx,t+\Delta t|\vecx,t)&=&\frac{1}{\sqrt{\det[4\pi\vecD\Delta t]}}\exp\left\{-\frac{\vecD^{-1}}{4\Delta t}:[\Delta\vecx-\veca\Delta t-\boldsymbol\tau\Delta t^2][\Delta\vecx-\veca\Delta t-\boldsymbol\tau\Delta t^2]\right\}\nonumber\\
    &\times&\left[1+D^{\rho\sigma}\nabla^\rho D^{\mu\nu}H_3^{\mu\nu\sigma}(\Delta\vecx,2\vecD\Delta t)\Delta t^2+\tilde\psi_{2}^{\mu\nu}H_2^{\mu\nu}(\Delta\vecx,2\vecD\Delta t)\Delta t^2\right.\nonumber\\
    &+&\left.\psi_{4}^{\mu\nu\sigma\rho}H_4^{\mu\nu\sigma\rho}(\Delta\vecx,2\vecD\Delta t)\Delta t^3+\psi_{6}^{\mu\nu\sigma\rho\alpha\beta}H_6^{\mu\nu\sigma\rho\alpha\beta}(\Delta\vecx,2\vecD\Delta t)\Delta t^4+\calo(\Delta t^{3/2})\right]\,.\label{eq:prop3/2}
\end{eqnarray}
where
\begin{equation}
\tilde\psi_{2}^{\mu\nu}=D^{\mu\alpha}D^{\nu\sigma}\nabla^\sigma[(\vecD^{-1})^{\alpha\rho}a^\rho]+\frac{1}{2}\frac{\partial D^{\mu\nu}}{\partial t}+\frac{1}{2}D^{\sigma\rho}\nabla^{\sigma}\nabla^{\rho}D^{\mu\nu} \,.\label{eq:psi2tilde}
\end{equation}

Equation~\eqref{eq:prop3/2} has the overshot order-$\Delta t^2$ term in the mean of the Gaussian, which is included for proper order-$\Delta t^{3/2}$ strong convergence. However, for the purpose of computing the entropy production (and other first-order derivatives of the trajectory) we wish to expand the propagator itself to order $\Delta t$. Thus, we shall drop $\boldsymbol\tau\Delta t^2$, and obtain an expression of the form suggested in Eq.~\eqref{eq:propSTRUC}, where $\boldsymbol\Phi$ is given in Eq.~\eqref{eq:phi}, and 
\begin{eqnarray}
    \Psi(\vecK|\vecx,t)&=&\frac14\tilde\psi_{2}^{\mu'\nu'}(\vecD^{-1})^{\mu'\mu}(\vecD^{-1})^{\nu'\nu}(K^{\mu\nu}-2D^{\mu\nu})\nonumber\\
    &+&\frac{1}{16}\psi_{4}^{\mu'\nu'\sigma'\rho'}(\vecD^{-1})^{\mu'\mu}(\vecD^{-1})^{\nu'\nu}(\vecD^{-1})^{\sigma'\sigma}(\vecD^{-1})^{\rho'\rho}\nonumber\\&&\times(K^{\mu\nu}K^{\sigma\rho}-2K^{\mu\nu}D^{\sigma\rho}-4K^{\mu\sigma}D^{\nu\rho}-4K^{\mu\rho}D^{\nu\sigma}-2K^{\sigma\rho}D^{\mu\nu}-4D^{\mu\nu}D^{\sigma\rho}-8D^{\mu\sigma}D^{\nu\rho})\nonumber\\
    &+&\psi_{6}^{\mu\nu\sigma\rho\alpha\beta}H_6^{\mu\nu\sigma\rho\alpha\beta}(\Delta\vecx,2\vecD\Delta t)\Delta t^3\,,\label{eq:psi}
\end{eqnarray}
where all the coefficients $\veca$, $\vecD$, $\boldsymbol\psi_2$, $\boldsymbol\psi_4$, and $\boldsymbol\psi_6$ were computed at $(\vecx,t)$, we utilized $\psi^{\mu\nu\sigma\rho}=\psi^{\nu\mu\sigma\rho}$ (without any other index permutation symmetry), and we have kept the the sixth Hermite polynomial implicit for compactness.
This demonstrates the central contribution of this paper\,---\,the ability to write a propagator of any order in a consistent fashion. Indeed, we see that the propagator (to order $\Delta t$) takes the form of Eq.~\eqref{eq:propSTRUC}. The calculation for higher-order propagators will become increasingly cumbersome. 

We conclude this section with a technical remark. The approximations Eqs.~\eqref{eq:prop1herm} and \eqref{eq:prop1} for order $\Delta t^{1/2}$, or \eqref{eq:prop3/2herm} and~\eqref{eq:prop3/2} for order $\Delta t$, provide an explicit expression for the propagator as a function of $\Delta\vecx$. They may be useful for computing short-time functionals of the propagator. However, Eqs.~\eqref{eq:prop1der} and~\eqref{eq:prop3/2der}, which include  derivatives, may be more convenient when computing moments. This is because, upon integration by parts, one is left with an average of lower-order polynomials with respect to the leading-order propagator, $P_{1/2}$ of Eq.~\eqref{eq:propSIMP}.

\subsection{Entropy production}\label{sec:EPR}

We proceed to compute the entropy production [Eq.~\eqref{eq:EPR}] for the Langevin equation and the related Fokker-Planck equation of Sec.~\ref{sec:model}. 
For that purpose, Eq.~\eqref{eq:infoheat} requires knowing the propagator in the ``reversed'' process, which is $P(x,\lambda|x+\Delta x,\lambda+\Delta \lambda)$.
In this expression, we brought back the protocol $\lambda$ (instead of just $t$) so the trajectory reversal will be clearer. 
To highlight the importance of the corrections presented here, we start with the na\"ive leading-order ($\Delta t^{1/2}$) propagator, Eq.~\eqref{eq:propSIMP}, 
\begin{multline}
    P_{1/2}(\vecx,\lambda|\vecx+\Delta\vecx,\lambda+\Delta \lambda)=\frac{1}{\sqrt{\det[4\pi\vecD(\vecx+\Delta\vecx,\lambda+\Delta \lambda)\Delta t]}}\\\times\exp\left\{-\frac{\vecD^{-1}(\vecx+\Delta\vecx,\lambda+\Delta \lambda)}{4\Delta t}:[\Delta\vecx+\veca(\vecx+\Delta\vecx,\lambda+\Delta \lambda)\Delta t][\Delta\vecx+\veca(\vecx+\Delta\vecx,\lambda+\Delta \lambda)\Delta t]\right\}\,.\label{eq:propSIMPrev}
\end{multline}
Essentially, we replaced $(\vecx,\lambda)$ with $(\vecx+\Delta\vecx,\lambda+\Delta\lambda)$ as the points where all the coefficients are evaluated (as the latter is now the initial point for the reversed process), and we replaced $\Delta\vecx$ with $-\Delta\vecx$ (as we need a displacement $-\Delta\vecx$ to arrive from $\vecx+\Delta\vecx$ to $\vecx$). With this, we revert back to $\lambda=t$.

We now compute the ratio of normalizations,
\begin{eqnarray}
    \frac12\ln\frac{\det[\vecD(\vecx+\Delta\vecx,t+\Delta t)]}{\det[\vecD(\vecx,t)]}&=&-\frac12D^{\mu\nu}(\vecx+\Delta\vecx/2,t)\nabla^\sigma(\vecD^{-1})^{\mu\nu}(\vecx+\Delta\vecx/2,t)\Delta x^\sigma\nonumber\\&-&\frac12D^{\mu\nu}(\vecx,t)\frac{\partial(\vecD^{-1})^{\mu\nu}(\vecx,t)}{\partial t}\Delta t+\calo(\Delta t^{3/2})\,.\label{eq:logD}
\end{eqnarray}
In Eq.~\eqref{eq:logD}, the term in the first line is evaluated at the midpoint following the Stratonovich chain rule. This means that partial derivatives with respect to $\vecx$ (at the midpoint) are taken using the standard chain rule, $\d\ln[\det(\vecD)]=\mathrm{tr}[\ln(\vecD^{-1}\circ\d\vecD)]$ (which is the Jacobi formula) and $\vecD^{-1}\circ\d\vecD=-\d(\vecD^{-1})\circ\vecD$. As preparation for later, we kept $[\cdot](\vecx+\Delta\vecx/2,t)=[\cdot](\vecx,t)+(1/2)\Delta\vecx\cdot\grad[\cdot](\vecx,t)+\calo(\Delta t)$; we will repeat this replacement in the remaining contributions to the entropy production as well. The term in the second line of Eq.~\eqref{eq:logD} is computed at the initial point since it is already of order $\Delta t$. 
Using Eq.~\eqref{eq:logD} we can compute the log-ratio of the leading-order propagators,
\begin{eqnarray}
    \ln\frac{P_{1/2}(\vecx+\Delta\vecx,t+\Delta t|\vecx,t)}{P_{1/2}(\vecx,t|\vecx+\Delta\vecx,t+\Delta t)}&=&
    \frac{1}{4}\nabla^{\sigma}(\vecD^{-1})^{\mu\nu}(\vecx+\Delta\vecx/2,t)\left[\frac{\Delta x^{\mu}\Delta x^{\nu}\Delta x^{\sigma}}{\Delta t}-2D^{\mu\nu}(\vecx+\Delta\vecx/2,t)\Delta x^{\sigma}\right]\nonumber\\
    &+&\frac{1}{4}\frac{\partial(\vecD^{-1})^{\mu\nu}(\vecx,t)}{\partial t}\left[\Delta x^{\mu}\Delta x^{\nu}-2D^{\mu\nu}(\vecx,t)\Delta t
    \right]\nonumber\\
    &+&\veca(\vecx+\Delta\vecx/2,t)\cdot\vecD^{-1}(\vecx+\Delta\vecx/2,t)\cdot\Delta\vecx+\calo(\Delta t^{3/2})\,.\label{eq:EPR1/2}
\end{eqnarray}
Note again the different points at which we computed the coefficients. The leading order should have been $\Delta\vecx\Delta\vecx/\Delta t\sim1$ (the scaling of pure Brownian motion) but it has cancelled out. Hence the reference to these quantities as ``first derivatives of trajectories''. Then, the division by $\Delta t$ within the exponent brings out the higher orders. We eventually wish to take the limit $\Delta t\to0$, meaning that the fluctuations in the second row are negligible and we may replace it by its average~\cite{book:schuss}, which is zero. No such operation can be done for the rest of the terms, which are of order $\Delta t^{1/2}$. Additionally, the term $\Delta\vecx\Delta\vecx\Delta\vecx$ cannot be replaced with another term of them form $[\cdot]\Delta t+[\cdot]\Delta\vecx$ and even more so to be brought to the form of the order-$\Delta t^{3/2}$ multiple stochastic integrals (see Sec.~\ref{sec:discmult}). Hence, one cannot write the above as a continuous-time stochastic differential quantity.

We now utilize the corrections we have found in this paper, of the form of Eq.~\eqref{eq:propSTRUC}, to correct the above expression for the entropy production. These additional contributions are seen in Eq.~\eqref{eq:insertinfoheat}. We know that the propagator is expanded as Eq.~\eqref{eq:propSTRUC}; \eg repeating Eq.~\eqref{eq:phi},
\begin{equation}
    \Delta\vecx\cdot\boldsymbol\Phi\left(\left.\frac{\Delta\vecx\Delta\vecx}{\Delta t}\right|\vecx,\lambda\right)=\frac18\nabla^\sigma (\vecD^{-1})^{\mu\nu}(\vecx,\lambda)\left[ 2D^{\mu\nu}(\vecx,\lambda)\Delta x^{\sigma}+4D^{\nu\sigma}(\vecx,\lambda)\Delta x^{\mu}-\frac{\Delta x^{\mu}\Delta x^{\nu}\Delta x^{\sigma}}{\Delta t}\right].\label{eq:dxPhi}
\end{equation}
with the explicit protocol value $\lambda$. Namely, if in Eq.~\eqref{eq:propSTRUC} we evaluated the coefficients within the polynomials $\boldsymbol\Phi$ and $\Psi$ at $(\vecx,t)$ [that is, at $(\vecx,\lambda)$], the propagator to obtain the reversed trajectory $\vecx+\Delta\vecx\to\vecx$ under $\lambda+\Delta\lambda\to\lambda$ reads
\begin{multline}
    P(\vecx,\lambda|\vecx+\Delta\vecx,\lambda+\Delta \lambda)=P_{1/2}(\vecx,\lambda|\vecx+\Delta\vecx,\lambda+\Delta \lambda)\\\times\left[1-\Delta\vecx\cdot\boldsymbol\Phi\left(\left.\frac{\Delta\vecx\Delta\vecx}{\Delta t}\right|\vecx+\Delta\vecx,\lambda+\Delta \lambda\right)+\Delta t\Psi\left(\left.\frac{\Delta\vecx\Delta\vecx}{\Delta t}\right|\vecx+\Delta\vecx,\lambda+\Delta \lambda\right)+\calo(\Delta t^{3/2})\right],\label{eq:propSTRUCrev}
\end{multline}
where the leading order reversed propagator is given in Eq.~\eqref{eq:propSIMPrev}.
In Eq.~\eqref{eq:insertinfoheat}, we have included all the terms that must be included. However, note that upon computing the log-ratio of the corrections and using the tools we presented throughout the paper,
\begin{multline}
\ln\frac{1+\Delta\vecx\cdot\boldsymbol\Phi(\Delta\vecx\Delta\vecx/\Delta t|\vecx,t)+\Delta t\Psi(\Delta\vecx\Delta\vecx/\Delta t|\vecx,t+\Delta t)+\calo(\Delta t^{3/2})}{1-\Delta\vecx\cdot\boldsymbol\Phi(\Delta\vecx\Delta\vecx/\Delta t|\vecx+\Delta\vecx,t+\Delta t)+\Delta t\Psi(\Delta\vecx\Delta\vecx/\Delta t|\vecx+\Delta\vecx,t+\Delta t)+\calo(\Delta t^{3/2})}\\=2\Delta\vecx\cdot\boldsymbol\Phi\left(\left.\frac{\Delta\vecx\Delta\vecx}{\Delta t}\right|\vecx+\frac{\Delta\vecx}{2},t\right)+\calo(\Delta t^{3/2}),\label{eq:EPRcorr}
\end{multline}
all the order-$\Delta t$ correction terms of Eq.~\eqref{eq:prop3/2} cancel out to order $\Delta t$ [namely, the distinction among evaluating them at $(\vecx+\Delta\vecx,\lambda+\Delta\lambda)$ instead of $(\vecx,\lambda)$ is of negligibly-high order]. Thus, it appears \textit{a posteriori} that the Milstein propagator [Eq.~\eqref{eq:prop1}] suffices for the calculation of entropy production.

Finally, we find the log-ratio of the propagators by inserting Eq.~\eqref{eq:dxPhi} in Eq.~\eqref{eq:EPRcorr}, and using  $\vecD^{-1}\cdot\d\vecD=-\d(\vecD^{-1})\cdot\vecD$, and  Eq.~\eqref{eq:EPR1/2}:
\begin{equation}
    \ln\frac{P(\vecx+\Delta\vecx,t+\Delta t|\vecx,t)}{P(\vecx,t|\vecx+\Delta\vecx,t+\Delta t)}=[\veca(\vecx+\Delta\vecx/2,t)-\grad\cdot\vecD(\vecx+\Delta\vecx/2,t)]\cdot\vecD^{-1}(\vecx+\Delta\vecx/2,t)\cdot\Delta\vecx+\calo(\Delta t^{3/2})\,.\label{eq:EPR3/2}
\end{equation}
Here, $\Delta\vecx\Delta\vecx\Delta\vecx$ has reassuringly cancelled out. 
Surprisingly, and due to the specific symmetry of the forward and reverse paths, the log-ratio of the propagators is just the result of Euler-Maruyama forward and backward propagators evaluated at the mid-point (Stratonovitch). This is a common practice in calculating the entropy production within the current literature (e.g., Refs.~\cite{CatesEnt22,CugliandoloJPA2017}). Importantly, although it holds for the entropy production, this practice is not general, and the propagators used are not calculated to a sufficient order in $\Delta t$. We discuss this point in detail in Sec.~\ref{app:equiv}.

Now that the calculation has been completed and all order-$\Delta t^{1/2}$ and -$\Delta t$ terms were included, we may take the limit $\Delta t\to0$. Since $\Delta\vecx\Delta\vecx\Delta\vecx$ and its analogoues have cancelled out, we see that we can make Eq.~\eqref{eq:EPR3/2} a continuous-time stochastic quantity. With the definitions of Sec.~\ref{sec:dilemma}, the informatic heat [Eq.~\eqref{eq:infoheat}] obeys the equation
\begin{equation}
    \Omega_t=\int_0^t\{[\veca(\vecX_{t'},t')-\grad\cdot\vecD(\vecX_{t'},t')]\cdot\vecD^{-1}(\vecX_{t'},t')\}\circ\d\vecX_{t'}\,.\label{eq:infoheatRES}
\end{equation}
We continue by examining the physical scenarios presented in Secs.~\ref{sec:FPE} and~\ref{sec:langevin} where the drift is $\veca=\bmu\cdot\vecf+\grad\cdot\vecD$ (see  Eqs.~\eqref{eq:diffeq3} and~\eqref{eq:SDE3}), such that  
\begin{equation}
    \Omega_t=\int_0^t\{[\bmu(\vecX_{t'},t')\cdot\vecf(\vecX_{t'},t')]\cdot\vecD^{-1}(\vecX_{t'},t')\}\circ\d\vecX_{t'}\,.
\end{equation}
When Einstein relation is valid $\vecD=k_\mathrm{B}T\bmu$, the informatic heat coincides with the heat~\cite{SorkinPR24,CatesEnt22},
\begin{equation}
    -k_\mathrm{B}T\Omega_t=-\int_0^t\vecf(\vecX_{t'},t')\circ\d\vecX_{t'}\,.
\end{equation}
In particular for conservative forces $\vecf(\vecX_{t'},t')=-\grad H(\vecX_{t'},t')$, using the stochastic chain rule~\cite{book:oksendal}, the informatic heat coincides with the change in energy along a trajectory~\cite{CatesEnt22},
\begin{equation}
    -k_\mathrm{B}T\Omega_t=H(\vecX_{t},t)-H(\vecX_{0},0)\,.
\end{equation}

Finally, for completeness, we compute the expression for the entropy production, $\Sigma_t$. First, we compute the change in the instantaneous stochastic entropy, $S_t=-\ln p(\vecX_t,t)$, which also appears in Eq.~\eqref{eq:infoheat}. This is a function of the microstate, and hence its change can be computed explicitly using the stochastic chain rule~\cite{book:oksendal},
\begin{equation}
    \d S_t=-\frac{\grad p(\vecX_t,t)}{p(\vecX_t,t)}\circ\d\vecX_t-\frac{1}{p(\vecX_t,t)}\frac{\partial p(\vecX_t,t)}{\partial t}\d t \,.
\end{equation}
In the following expressions for $\d S_t$, $\d\Omega_t$, and $\d\Sigma_t$ we will not write $(\vecX_t,t)$ for brevity. Converting the Stratonovich product into It\^o product~\cite{book:schuss}, $\mathbf{g}(\vecX_t,t)\circ\d\vecX_t=\mathbf{g}(\vecX_t,t)\cdot\d\vecX_t+\vecD:\grad\mathbf{g}(\vecX_t,t)\d t$, and using Eqs.~\eqref{eq:SDE} and \eqref{eq:FPE} to express $\d\vecX_t$ and $\partial p/\partial t$, respectively, we arrive at
\begin{equation}
    \d S_t=\left[\grad\cdot\left(\frac{\vecJ}{p}\right)-\frac{1}{p}\grad(\vecD\grad p)\right]\d t-\left[\frac{\grad p}{p}\cdot\vecb\right]\cdot\d\vecW_t\,,
\end{equation}
where $\vecJ(\vecx,t)$ is the probability flux of the Fokker-Planck equation, Eq.~\eqref{eq:FPEflux}. Converting the Stratonovich product into It\^o in Eq.~\eqref{eq:infoheatRES} and using Eq.~\eqref{eq:SDE}, we get an analogous expression for the informatic heat 
\begin{equation}
    \d\Omega_t=[(\veca-\grad\cdot\vecD)\cdot\vecD^{-1}\cdot(\veca-\grad\cdot\vecD)+\grad\cdot(\veca-\grad\cdot\vecD)]\d t+[(\veca-\grad\cdot\vecD)\cdot\vecD^{-1}\cdot\vecb]\cdot\d\vecW_t\,.
\end{equation}
Combining both equations, we find the entropy production of the process of Sec.~\ref{sec:langevin},
\begin{equation}
    \d\Sigma_t=\left[\frac{\vecJ}{p}\cdot\vecD^{-1}\cdot\frac{\vecJ}{p}+\frac{2}{p}\grad\cdot\vecJ\right]\d t+\left[\frac{\vecJ}{p}\cdot\vecD^{-1}\cdot\vecb\right]\cdot\d\vecW_t\,.\label{eq:dSIGMAlangevin}
\end{equation}
Using probability conservation [$\int\d\vecx\grad\cdot\vecJ(\vecx,t)=\oint\d\delta\vecx\cdot\vecJ(\vecx,t)=0$], we get that the entropy production rate averaged over noise ($\langle\d\vecW_t\rangle_\vecW=\veco$) and ensemble ($\langle[\cdot](\vecX_t,t)\rangle_\vecx=\int \d\vecx p(\vecx,t)[\cdot](\vecx,t)$) is a nonnegative quadratic form~\cite{SeifertPRL05}, 
\begin{equation}
\langle\dot\Sigma_t\rangle_{\vecW,\vecx}=\int\d\vecx \frac{\vecJ(\vecx,t)\cdot\vecD^{-1}(\vecx,t)\cdot\vecJ(\vecx,t)}{p(\vecx,t)}\,.
\end{equation}
This must be the case since the average entropy production [Eq.~\eqref{eq:EPR}] is the nonnegative Kullback-Leibler divergence. 

With that, we have demonstrated the underlying tools for computing consistently the time-reversal symmetry breaking (the entropy production) of arbitrary Fokker-Planck equations. In fact, upon taking the ensemble average of Eq.~\eqref{eq:infoheatRES} and transitioning from the stochastic representation $\int\langle[\cdot](\vecX_t,t)\circ\d\vecX_t\rangle$ to the probabilistic one $\int[\cdot](\vecx,t)\vecJ(\vecx,t)\d\vecx$, we reassuringly arrive at an identical result to Ref.~\cite{kapplerARX24}.


\section{Non-It\^o propagators} \label{app:equiv}

In this section, we revisit the SDE in a general convention, Eq.~\eqref{eq:langevin_alpha}. The transition from Eq.~\eqref{eq:langevin} to the simplest Euler-Maruyama approximation, Eq.~\eqref{eq:dXnaive}, motivates~\cite{LubenskyPRE07} a similar approximation for the general Eq.~\eqref{eq:langevin_alpha}. The Euler-Maruyama expansion for the $\alpha$ convention reads
\begin{equation}
    \Delta\vecX_{t} \simeq \veca_\alpha(\vecX_{t}+\alpha\Delta \vecX_t,\Lambda_t+\alpha\Delta \Lambda_{t})\Delta t + \vecb(\vecX_{t}+\alpha\Delta \vecX_t,\Lambda_t+\alpha\Delta \Lambda_{t})\cdot\Delta\vecW_{t}\,.\label{eq:dXt_alpha_apprx}
\end{equation}
As in Eq.~\eqref{eq:langevin_alpha}, the drift here is $a^\mu_\alpha=a^\mu-\alpha b^{\sigma\nu}\nabla^\sigma b^{\mu\nu}$. Observe that the coefficients in this Euler-Maruyama expression are evaluated at an intermediate point as dictated by the convention, $(\vecX_{t+\alpha\Delta t},\Lambda_{t+\alpha\Delta t})=(\vecX_{t}+\alpha\Delta \vecX_t,\Lambda_t+\alpha\Delta \Lambda_{t})$. (For example, It\^o convention $\alpha=0$ evaluates the coefficient at the start point, while the Stratonovich convention $\alpha=1/2$ at the midpoint.)

\subsection{Entropy production}

It has been argued in Ref.~\cite{CatesEnt22} that the leading-order (Gaussian) propagator suffices for the calculation of the informatic heat, so long as it is ``correctly'' discretized. Namely, one must express the backward propagator in the complementary convention to the one of the forward propagator,
\begin{equation}
    \dot\Omega_t=\lim_{\Delta t\to0}\frac{1}{\Delta t}\ln\frac{P^\alpha_{1/2}(\vecx+\Delta\vecx,\lambda+\Delta\lambda|\vecx,\lambda)}{P^{1-\alpha}_{1/2}(\vecx,\lambda|\vecx+\Delta\vecx,\lambda+\Delta\lambda)}\,.
\end{equation}
That is, the $\alpha$ convention is used in the enumerator, in contrast to $1-\alpha$ of the denominator. (We write $P^\alpha_{1/2}(\vecx+\Delta\vecx,\lambda+\Delta\lambda|\vecx,\lambda)$ explicitly in Eq.~\eqref{eq:prop_alpha} below.)
The motivation for this discretization stems from the observation that using complementary conventions means that the random Wiener noise is sampled at the same points in both directions~\cite{CatesEnt22,LubenskyPRE07}. While intuitive, there is no mathematical basis why this is a sufficient or even necessary condition for ``correct'' computations.
Moreover, as discussed in Sec.~\ref{sec:dilemma}, \emph{this cannot be true}: All $\alpha$ conventions represent the same stochastic process, and hence must give identical statistics and FPE, and in particular all propagators must be identical. If one takes the limit $\Delta t\to0$ consistently, there must not be an explicit dependence on the discretization method $\alpha$. 

Since a ``wrong'' result would have been obtained for non-complementary conventions for the forward ($\alpha$) and backward (inverted, $1-\alpha$) propagators, this suggests that the small-$\Delta t$ expansion of Eq.~\eqref{eq:dXt_alpha_apprx} was not kept to sufficiently-high order. In other words, terms that would have cancelled convention dependent corrections are absent. This is evident in Eq.~\eqref{eq:insertinfoheat}\,---\,as we established there, the propagator must include the next two orders beyond the leading order Eq.~\eqref{eq:dXt_alpha_apprx}.
In contrast, our derivation of Sec.~\ref{sec:EPR}, does not depend on the choice of discretization, rather, we used stochastic Taylor expansions to compute the corrections from the discrepancy function, Eq.~\eqref{eq:error}. For simplicity, both the forward and backward propagators were computed from the It\^o representation.

In this section, we present an extensive comparison with previous methods for evaluating the propagators~\cite{LubenskyPRE07,CugliandoloJPA2017,CugliandoloJPA19,CugliandoloREV23} and their application to entropy production~\cite{book:stochastic,CatesEnt22}. Detailed derivations are deferred to Appendices~\ref{sec:nonitostoc} and~\ref{sec:moment}. The goal is to resolve previous technical inconsistencies and explain why the calculation of the entropy production in, \eg Ref.~\cite{CatesEnt22} turns out to be correct. We further exemplify ``toy functionals'' for which a complementary choice of propagator conventions does not apply.

\subsection{A comment on path-integral methods}

We have presented the It\^o Euler-Maruyama propagator in Eq.~\eqref{eq:propSIMP}.
For time-independent Langevin processes (\ie $t$-independent $\veca$ and $\vecb$), the non-It\^o Euler-Maruyama propagator can be found in the literature~\cite{LubenskyPRE07,CatesEnt22,CugliandoloJPA2017},
\begin{equation}
    P_{1/2}^\alpha(\vecx+\Delta\vecx,t+\Delta t|\vecx,t)=\frac{1}{\sqrt{(4\pi\Delta t)^d\det\vecD(\vecx+\alpha\Delta\vecx)}}\exp[-\lag^\alpha(\vecx+\Delta\vecx;\vecx)\Delta t] \,.\label{eq:prop_alpha}
\end{equation}
Here, $\lag^\alpha$ is often called the ``dynamical Lagrangian'' 
for the $\alpha$th convention,
\begin{eqnarray}
    \nonumber\lag^\alpha(\vecx+\Delta\vecx;\vecx)&=&
    \left\{\frac{\Delta\vecx}{\Delta t}-[\veca_\alpha-\alpha\vecb\cdot(\grad\cdot\vecb)](\vecx+\alpha\Delta\vecx)\right\}
    \cdot\frac{\vecD^{-1}(\vecx+\alpha\Delta\vecx)}{4}\cdot
    \left\{\frac{\Delta\vecx}{\Delta t}-[\veca_\alpha-\alpha\vecb\cdot(\grad\cdot\vecb)](\vecx+\alpha\Delta\vecx)\right\}
    \\
    &+&\alpha\grad\cdot\veca_\alpha(\vecx+\alpha\Delta\vecx)+\frac{\alpha^2}{2}\left[(\nabla^\nu b^{\mu\sigma})(\nabla^\mu b^{\nu\sigma})-(\nabla^\mu b^{\mu\sigma})(\nabla^\nu b^{\nu\sigma})\right](\vecx+\alpha\Delta\vecx)\,,\label{eq:lagrangian}
\end{eqnarray}
which depends on the displacement $\Delta\vecx$, and all functions within it are evaluated at $\vecx+\alpha\Delta\vecx$. This propagator can then be used to, \eg obtain the Fokker-Planck equation or compute moments~\cite{LubenskyPRE07,CugliandoloREV23,book:schuss}. With the dynamic action in hand, using the Markov property of Eq.~\eqref{eq:langevin}, it is common to represent the probability distribution to obtain a  trajectory $\left(\vecx_0,\vecx_1,\ldots,\vecx_M\right)$ (with $T=M \Delta t$), for example, as appearing in Eq.~\eqref{eq:EPR}, using the ``dynamical action''~\cite{LubenskyPRE07,book:stochastic},
\begin{eqnarray}
    \Pr[\overrightarrow\vecx]
    &=& Z^{-1} p(\vecx_0,0)\exp\left[-\int_0^{t_\mathrm{final}}\d t\,\lag^\alpha(\vecx+\d\vecx;\vecx)\right] \nonumber \\
    &\sim& p(x_0,0)\prod_{i=1}^{M}P_{1/2}^\alpha(\vecx_i,i\Delta t|\vecx_{i-1},(i-1)\Delta t) \,,
\end{eqnarray}
where $Z$ is a normalization constant.

The notation above facilitates adaptation of tools from quantum mechanics to treat diffusion processes~\cite{LubenskyPRE07}. 
The path integral representation is possible by identifying $\Delta\vecx/\Delta t$ as $\dot\vecx$, so $\Delta\vecx$ can be converted to $\dot\vecx\d t$, and $\Delta\vecx\Delta\vecx/\Delta t$ to $\dot\vecx\dot\vecx\d t$~\cite{LubenskyPRE07}.
We remark that due to terms such as $\Delta\vecx\Delta\vecx\Delta\vecx/\Delta t$, a path integral representation of the exact high-order Eq.~\eqref{eq:prop3/2herm} [and also Eq.~\eqref{eq:prop1herm}] is impossible, as there is just a single integration over time. We elaborate on this point in Sec.~\ref{sec:summary}.

%

\subsection{Toy functionals}\label{sec:toy}

In order to demonstrate that a combination of $\alpha$ and $1-\alpha$ conventions does not always work, we consider a toy functional, which is a first derivative of the trajectory.
For simplicity, consider the case in which Einstein relation holds, $\beta\vecI=\bmu\cdot\vecD^{-1}$, and that all forces are conservative central forces, $\vecf(\vecx)=-\grad H(|\vecx|)$. Then, the steady-state distribution is given by the isotropic Boltzmann factor $p(\vecx,t\to\infty)\sim e^{-\beta H(\vecx)}$. Let us assume the particle of interest is charged, and that one applies a concentric magnetic field. This will not change the steady state, $p(\vecx,t\to\infty)\sim e^{-\beta H(\vecx)}$, as a magnetic field will only lead to rotations of the charged particles around the center. It is, therefore, impossible to distinguish among the two systems via the steady-state distribution. However, the underlying dynamics are clearly different. A natural way to quantify the difference between the two dynamics is to compute the Kullback-Liebler divergence from the statistics of trajectories with the magnetic field to the one without it. Accordingly, we have a motivation to calculate the following ``toy functional''
\begin{equation}
    \dot\Pi_t^{12}=\lim_{\Delta t\to0}\frac{1}{\Delta t}\ln\frac{P^1(\vecx+\Delta\vecx,t+\Delta t|\vecx,t)}{P^2(\vecx+\Delta\vecx,t+\Delta t|\vecx,t)}\,, \label{eq:toy}
\end{equation}
where $P^i(\vecx+\Delta\vecx,t+\Delta t|\vecx,t)$ denotes the propagator with drift $\veca^i$.
Eq.~\eqref{eq:toy} is of the form we previously referred to as a first derivative of trajectories, because it is the difference between two functions of the trajectory (here propagators) that differ by order $\Delta t$, then re-scaled to order $1$ via a division by $\Delta t$. Thus, we should utilize the order-$3/2$ propagator.
This toy functional detects the difference in biases that resulted from different drifts that are imposed upon a system. 

We continue by computing the log ratio among the propagator of Eq.~\eqref{eq:langevin} with two different drift terms, $\veca^1$ and $\veca^2$. Each of these propagators are given by Eq.~\eqref{eq:prop3/2}, with the appropriate drift. Namely, 
\begin{eqnarray}
    \nonumber P^i(\vecx+\Delta\vecx,t+\Delta t|\vecx,t)&=&P^i_{1/2}\{\vecx+\Delta\vecx,t+\Delta t|\vecx,t)\left[1+D^{\rho\sigma}\nabla^\rho D^{\mu\nu}H_3^{\mu\nu\sigma}(\Delta\vecx,2\vecD\Delta t)\Delta t^2+\tilde{\psi}{}_{2}^{i,\mu\nu}H_2^{\mu\nu}(\Delta\vecx,2\vecD\Delta t)\Delta t^2\right.
    \\&+&\left.\psi_{4}^{\mu\nu\sigma\rho}H_4^{\mu\nu\sigma\rho}(\Delta\vecx,2\vecD\Delta t)\Delta t^3+\psi_{6}^{\mu\nu\sigma\rho\zeta\xi}H_6^{\mu\nu\sigma\rho\zeta\xi}(\Delta\vecx,2\vecD\Delta t)\Delta t^4+\calo(\Delta t^{3/2})\right\}\,,\label{eq:prop3/2zero}
\end{eqnarray}
where
\begin{equation}
    P^i_{1/2}(\vecx+\Delta\vecx,t+\Delta t|\vecx,t)=\frac{1}{\sqrt{\det[4\pi\vecD(\vecx,t)\Delta t]}}\exp\left\{-\frac{\vecD^{-1}(\vecx,t)}{4\Delta t}:[\Delta\vecx-\veca^i(\vecx,t)\Delta t][\Delta\vecx-\veca^i(\vecx,t)\Delta t]\right\}\,,
\end{equation}
$\tilde{\boldsymbol\psi}{}^1_2$, $\tilde{\boldsymbol\psi}{}^2_2$ also depend on $\veca$ 
[see Eq.~\eqref{eq:psi2tilde}], while $D^{\rho\sigma}\nabla^\rho D^{\mu\nu}$, $\boldsymbol\psi_4$, and $\boldsymbol\psi_6$ do not depend on $\veca$ 
and hence are the same to both propagators.
Substituting into~\eqref{eq:toy}, to order-$\Delta t$ the log-ratio is
\begin{eqnarray}
    \nonumber \ln\frac{P^1(\vecx+\Delta\vecx,t+\Delta t|\vecx,t)}{P^2(\vecx+\Delta\vecx,t+\Delta t|\vecx,t)}&=&\frac{1}{4}\vecD^{-1}(\vecx,t):\left\{2\left[\veca^1(\vecx,t)-\veca^2(\vecx,t)\right]\Delta\vecx-\left[\veca^1(\vecx,t)\veca^1(\vecx,t)-\veca^2(\vecx,t)\veca^2(\vecx,t)\right]\Delta t\right\}\\
    &+&(\tilde\psi{}^{1,\mu\nu}_{2}-\tilde\psi{}_{2}^{2,\mu\nu})H_2^{\mu\nu}(\Delta\vecx,2\vecD\Delta t)\Delta t^2+\calo(\Delta t^{3/2})\,.
\end{eqnarray}
Since the last term is already of order $\Delta t$, the fluctuations in $\vecH_2(\Delta\vecx,2\vecD\Delta t)=\Delta\vecx\Delta\vecx-2\vecD(\vecx,t)\Delta t$ are negligible for $\Delta t\to0$. Thus, we can insert its mean, which is zero. Overall, we arrive at
\begin{eqnarray}    \nonumber\Pi_t&=&\frac12\int_0^t[\veca^1(\vecX_{t'},t')-\veca^2(\vecX_{t'},t')]\cdot\vecD^{-1}(\vecX_{t'},t')\cdot\d\vecX_{t'}\\
&-&\frac14\int_0^t\vecD^{-1}(\vecX_{t'},t'):[\veca^1(\vecX_{t'},t')\veca^1(\vecX_{t'},t')-\veca^2(\vecX_{t'},t')\veca^2(\vecX_{t'},t')]\d t'\,.\label{eq:toyresult}
\end{eqnarray}

Similar to the case with entropy production, it is possible to bypass the rigorous calculation above using similar reasoning as in Ref.~\cite{CatesEnt22}. Again, the order-$3/2$ terms vanished. We next need to ``engineer'' propagator conventions that account for the Milstein corrections. Computing entropy production required taking complementary conventions to recreate the symmetric part of the Milstein correction terms. For the toy example at hand, Eq.~\eqref{eq:toy}, we need to use the same convention for both processes. Namely, one can show that, 
\begin{equation}
    \dot\Pi_t=\lim_{\Delta t\to0}\frac{1}{\Delta t}\ln\frac{P^{1,\alpha}_{1/2}(\vecx+\Delta\vecx,\lambda+\Delta\lambda|\vecx,\lambda)}{P^{2,\alpha}_{1/2}(\vecx+\Delta\vecx,\lambda+\Delta\lambda|\vecx,\lambda)}\,.
\end{equation}
With this combination, the Milstein correction terms are accurately reproduced. Were we to take complementary conventions here, we would have reached a wrong result.

To further demonstrate, consider another illustrative functional, now with a different symmetry than the above two. To compute the lagged log-ratio of the form
\begin{equation}
\frac{1}{\Delta t}\ln \frac{P(\vecx+\Delta\vecx,\lambda+\Delta\lambda|\vecx,\lambda)}{P(\vecx+(1+\beta)\Delta\vecx,\lambda+(1+\beta)\Delta\lambda|\vecx+\beta\Delta\vecx,\lambda+\beta\Delta\lambda)} \, .
\end{equation}
an appropriate choice for a low-order convention-dependent propagators would have be

\begin{equation}
\frac{1}{\Delta t}\ln \frac{P^{\alpha+\beta}_{1/2}(\vecx+\Delta\vecx,\lambda+\Delta\lambda|\vecx,\lambda)}{P^{\alpha}_{1/2}(\vecx+(1+\beta)\Delta\vecx,\lambda+(1+\beta)\Delta\lambda|\vecx+\beta\Delta\vecx,\lambda+\beta\Delta\lambda)} \, .
\end{equation}
In contrast, for the following, more complicated functional, the low-order non-It\^o propagators cannot be determined \textit{a priori}:
\begin{equation}
    \frac{1}{\Delta t}\ln\frac{e^{aP(\vecx+\Delta\vecx,\lambda+\Delta\lambda|\vecx,\lambda)}-1}{aP(\vecx,\lambda|\vecx+\Delta\vecx,\lambda+\Delta\lambda)}\,.
\end{equation}
Here $a$ is some constant that fixes units. (For $a\to0$, it is exactly the informatic heat rate.)

With that, we notice that computing any functional of $P(\vecx+\Delta\vecx,t+\Delta t|\vecx,t)$ to order $\Delta t$ always leads to the disappearance of the order-$\Delta t$ terms in Eq.~\eqref{eq:propGEN}. Namely, suppose one computes the arbitrary functional $\Delta t^{-1}\mathcal F[P(\vecx+\Delta\vecx,t+\Delta t|\vecx,t)]$. We Taylor expand 
\begin{equation}
    \mathcal F[P]=\mathcal F[P_{1/2}]+\frac{\delta\mathcal F[P_{1/2}]}{\delta P}\Delta\vecx\cdot\boldsymbol\Phi+\frac{\delta\mathcal F[P_{1/2}]}{\delta P}\Delta t\Psi+\frac{1}{2}\frac{\delta^2\mathcal F[P_{1/2}]}{\delta P^2}(\Delta\vecx\cdot\boldsymbol\Phi)^2+\calo(\Delta t^{3/2}).
\end{equation}
Due to normalization, we know that the means of both $\Delta\vecx\cdot\boldsymbol\Phi$ and $\Delta t\Psi$ are necessarily zero. Since the former is of order $\Delta t^{1/2}$, we cannot replace it with its mean. 
However,  $\Delta t\Psi$ is already of order $\Delta t$ at least, so we replace it with its zero mean.
Similarly, $(\Delta\vecx\cdot\boldsymbol\Phi)^2$ can be replaced by its mean, which generally does not vanish (rather, it is the nonnegative variance of $\Delta\vecx\cdot\boldsymbol\Phi$).  Overall, we get,
\begin{equation}
    \mathcal F[P]=\mathcal F[P_{1/2}]+\frac{\delta\mathcal F[P_{1/2}]}{\delta P}\Delta\vecx\cdot\boldsymbol\Phi+\frac{1}{2}\frac{\delta^2\mathcal F[P_{1/2}]}{\delta P^2}\langle(\Delta\vecx\cdot\boldsymbol\Phi)^2|\vecX_t,t\rangle+\calo(\Delta t^{3/2})\,.
\end{equation}
Thus, the Milstein propagator suffices for computing first derivatives of the trajectory, so long as they only contain the propagator.
However, if one is after a more elaborate functionals, mixing the propagator and $\Delta\vecx$, the order-$\Delta t$ corrections of Eq.~\eqref{eq:prop3/2} would resurface. One such arbitrary example would be $(\Delta\vecx\Delta\vecx/\Delta t)\cdot\mathcal{F}[P(\vecx+\Delta\vecx,t+\Delta t|\vecx,t)]$, as now the mean of $\Delta\vecx\Delta\vecx\Psi(\Delta\vecx\Delta\vecx/\Delta t|\vecx,t)$ is not zero.

To conclude, when considering general functionals of first derivative forms, one can use, robustly, the Milstein propagator of Eq.~\eqref{eq:prop1}. If, instead, one insists upon using the convention-dependant Euler-Maruyama propagators of Ref.~\cite{CatesEnt22}, as before, one must carefully choose the conventions that cook up the correct Milstein corrections. The working choice can oftentimes be found by identifying the symmetries of the dynamics. This idea was demonstrated above for the entropy production and for other toy functionals.

\section{Summary}\label{sec:summary}

In this paper, we have studied the short-time evolution properties of an arbitrary diffusion process, given by either the overdamped Langevin equation, Eq.~\eqref{eq:langevin}, or its corresponding Fokker-Planck equation, Eq.~\eqref{eq:FPE}. We have proposed a consistent methodology for obtaining short-time expansions of the propagators. To leading order, the propagator is Gaussian. Higher order corrections, written in the form~\eqref{eq:propSTRUC}, are polynomials of the displacement, where the first two corrections were computed explicitly.\footnote{Reassuringly, in one dimension,  our Milstein and order-$\Delta t^{3/2}$ propagators agree respectively with the $K=1$ and $K=2$ propagators of Ref.~\cite{kapplerARX24}.}

The approximate propagator is used to compute the entropy production, Eq.~\eqref{eq:infoheat}\,---\,the extent to which a particular stochastic trajectory breaks time-reversal symmetry. It is a key quantity in modern nonequilibrium thermodynamics~\cite{book:stochastic}; with the Einstein relation, it is related to the thermodynamic heat dissipation ~\cite{SorkinPR24,CatesEnt22,PRX2021}. Reassuringly, our precise derivation has given the result that can be found in literature~\cite{CatesEnt22}, despite previous works' propagators being convention-dependent. With our rigorous derivation we explain how cancellation of errors leads to this result, and describe,  
via several toy functionals, how other functionals of the propagators can be evaluated, and what orders of expansions should be kept.

The consistent method with which we derived the propagator expansion pinpoints the mathematical inconsistencies that occurred in previous approaches to derive an analytical expression for the entropy production~\cite{LubenskyPRE07,CugliandoloJPA2017,CugliandoloJPA19,CugliandoloREV23,CatesEnt22,book:stochastic,OnsagerPRB53}. They originate from the apparent dependence on noise convention. Since the propagator and entropy production are continuous-time properties, they cannot depend on convention, discretization method, or representation of the physical stochastic process. Extensive technical comparisons were detailed in Sec.~\ref{app:equiv}. 
Since the propagator is not convention dependent, the leading order Euler-Maruyama propagator is just the It\^o one. 
We show how for specific cases (such as the calculation of the entropy production) meticulous construction of a convention-dependent Euler-Maruyama propagator that is chosen to satisfy particular symmetries, can give the correct answer. However, these constructions are not general. Instead, we suggest to resort to consistent derivations of the propagator to order $\Delta t$, which will give the correct result for any functional of the propagator.

We comment that the path integral approach~\cite{LubenskyPRE07} used, for example, to prove fluctuation theorems and compute entropy production, may occasionally also be inconsistent. While one is able to make the replacements $\sum\Delta\vecx\to\int\dot\vecx\d t$ and $\sum\Delta\vecx\Delta\vecx/\Delta t\to\int\dot\vecx\dot\vecx\d t$, there is no analogue for $\sum\Delta\vecx\Delta\vecx\Delta\vecx/\Delta t$. This term already appears in the first correction of the leading-order propagator [the Milstein propagator, Eq.~\eqref{eq:prop1}] whenever the diffusivity is position dependent. On one hand,  for computation of ``standard'' averages such as correlation and response functions~\cite{LubenskyPRE07,MartinPRE21}, or the entropy production of processes with position-independent diffusivity~\cite{MartinJSTAT21}, the path-integral approach and its consequential perturbation techniques are still useful since the leading-order propagator suffices.
%
%
On the other hand, converting the convolution of propagators into a path integral (as done, e.g., in Ref.~\cite{CatesEnt22}) in order to compute general first derivatives of the trajectory, such as the entropy production, is not rigorous. This is because, as we emphasized in this paper, the first two corrections for the propagator must be included in these cases. The inadequacy of the path integral approach suggests that the Markov decomposition of Eq.~\eqref{eq:infoheat} is a more consistent route for obtaining the entropy production accumulated over an entire path.

In some specific cases, for example the entropy production, the symmetries of the functional with respect to time reversal are known. Therefore, choosing an appropriate complementary, low-order non-It\^o conventions works as the Milstein corrections cancel out. This justifies the use of path integral approached in this specific case~\cite{LubenskyPRE07}. In general, for complicated functionals of the propagator, the path-integral approach may fail.


We suggest another possible application for the above expansions of the propagator as a high-order numerical scheme for SDEs. Since the multiple stochastic integrals of Sec.~\ref{sec:discmult} are all correlated and some are non-Gaussian, an implementation of high-order Brownian simulations of $\vecX_t$ by approximating Eq.~\eqref{eq:langevin} via Eq.~\eqref{eq:dX1} or Eq.~\eqref{eq:dX3/2} is not trivial~\cite{book:KloedenPlaten}. Namely, it require expressing $\Delta\vecY_t$, $\Delta\vecZ_t$, $\Delta\vecK_t$, and $\Delta\vecL_t$ in terms of infinitely-many uncorrelated normal-distributed random variables, which of course is impossible and, as an approximation, only finitely-many such variables are taken~\cite{book:KloedenPlaten}. Here, on the other hand, we find a simpler alternative to finding averages over Brownian variables based upon the importance sampling Monte-Carlo method~\cite{book:ISMC}. Suppose we wish to evaluate some function of the trajectory, e.g., the correlation function $\langle A(\vecX_t)B(\vecX_0)\rangle$ where $A$ and $B$ are some functions. We discretize this time duration into $M$ steps $\Delta t=t/M$. What the ensemble average implies is
\begin{equation}
    \langle A(\vecX_t)B(\vecX_0)\rangle=\int \d\vecx_0\int\d\vecx_1\cdots\int\d\vecx_Mp(\vecx_0)P(\vecx_1,\Delta t|\vecx_0,0)\cdots P(\vecx_M,M\Delta t|\vecx_{M-1},(M-1)t)A(\vecx_M)B(\vecx_0)\, ,
\end{equation}
where $p(\vecx_0)$ is the probability density of initial state, while $P(\vecx+\Delta\vecx,t+\Delta t|\vecx,t)$ is the short-time propagator to the desirable order. The latter is an increasingly-complicated distribution for higher-order propagators in small $\Delta t$, and hence it may not be possible to draw $\vecx_m-\vecx_{m-1}$ directly from it. However, by dividing and multiplying each propagator by the leading-order Gaussian propagator, $P_{1/2}(\vecx+\Delta\vecx,t+\Delta t|\vecx,t)$, we arrive at
\begin{eqnarray}
    \nonumber & &\langle A(\vecX_t)B(\vecX_0)\rangle=\int \d\vecx_0p(\vecx_0)\int\d\Delta\vecx_1\cdots\int\d\Delta\vecx_M 
    A(\vecx_0+\Delta\vecx_1+\cdots+\Delta\vecx_M)B(\vecx_0) \times
    \\
    \nonumber& &\times P_{1/2}(\vecx_0+\Delta\vecx_1,\Delta t|\vecx_0,0)\cdots P_{1/2}(\vecx_0+\Delta\vecx_1+\cdots+\Delta\vecx_M,M\Delta t|\vecx_0+\Delta\vecx_1+\cdots+\Delta\vecx_{M-1},(M-1)t) \\
    & & \times\frac{P(\vecx_0+\Delta\vecx_1,\Delta t|\vecx_0,0)}{P_{1/2}(\vecx_0+\Delta\vecx_1,\Delta t|\vecx_0,0)}\cdots \frac{P(\vecx_0+\Delta\vecx_1+\cdots+\Delta\vecx_M,M\Delta t|\vecx_0+\Delta\vecx_1+\cdots+\Delta\vecx_{M-1},(M-1)t)}{P_{1/2}(\vecx_0+\Delta\vecx_1+\cdots+\Delta\vecx_M,M\Delta t|\vecx_0+\Delta\vecx_1+\cdots+\Delta\vecx_{M-1},(M-1)t)}\,.\label{eq:path_simulation}
\end{eqnarray}

Equation~\eqref{eq:path_simulation} suggests the following simulation scheme. Draw each displacement from the leading-order Gaussian distribution in sequence (namely, first draw $\Delta\vecx_1$, then $\Delta\vecx_2$, etc.). It must be sequential since the Gaussian at the $m$th timestep has the mean $\veca(\vecx_m,m\Delta t)\Delta t$ [or $\veca(\vecx_m,m\Delta t)\Delta t+\boldsymbol\tau(\vecx_m,m\Delta t)\Delta t^2$ for the order-$\Delta t^{3/2}$ scheme, Eq.~\eqref{eq:prop3/2}] and variance $2\vecD(\vecx_m,m\Delta t)\Delta t$, which depend on the displacements that were drawn earlier. With this drawn $\{\Delta\vecx_m\}$, compute the average of $A(\vecx_0)B(\vecx_M)$ multiplied by the product of weight function $P(\vecx+\Delta\vecx,t+\Delta t|\vecx,t)/P_{1/2}(\vecx+\Delta\vecx,t+\Delta t|\vecx,t)=1+\Delta\vecx\cdot\boldsymbol\Phi(\Delta\vecx\Delta\vecx/\Delta t|\vecx,t)+\Delta t\Psi(\Delta\vecx\Delta\vecx/\Delta t|\vecx,t)+\cdots$ [truncated at the sought after order; Eqs.~\eqref{eq:prop1/2}, \eqref{eq:prop1}, or~\eqref{eq:prop3/2}] for every single step. This allows computing averages over noise realizations with just a single multi-dimensional normal-distributed variable $\Delta\vecX_t$ (instead of infinite copies) for every time step, by sampling $\Delta\vecX_t$ from a Gaussian and then rescaling it according to Eqs.~\eqref{eq:prop1/2}, \eqref{eq:prop1}, or~\eqref{eq:prop3/2}. They exactly correspond to the approximations appearing in Eqs.~\eqref{eq:dX1/2}, \eqref{eq:dX1}, and~\eqref{eq:dX3/2}. Note, however, that this proposed scheme only converges in a weak sense (that is, only on average).

We hope that the role of high-order propagator expansions in stochastic thermodynamics will be further studied, especially when considering functionals or expectation values that do not satisfy known special symmetries~\cite{kapplerARX24}.

\section*{Acknowledgements}
We thank Haim Diamant for extensive and helpful discussions. T.M. acknowledges funding from the Israel Science Foundation (Grant No. 1356/22).

\appendix

\section{Exactly-solvable example: Geometric Brownian Motion}\label{app:GBM}

We briefly demonstrate our results for geometric Brownian motion (GBM). GBM is a one-dimensional, exactly-solvable stochastic differential equation. It is particularly useful in mathematical finance, where it is used to model the time evolution of option prices in the Black–Scholes model~\cite{book:oksendal}. It is given by the It\^o SDE,
\begin{equation}
    \d X_t=\mu X_t\d t+\sigma X_t\cdot\d W_t\,,\label{eq:SDE_GBM}
\end{equation}
where, as usual, $W_t$ is the Wiener process. The equation is linear, but with multiplicative noise; the diffusion coefficient here is $D(x)=\sigma^2x^2/2$ and the drift is $a(x)=\mu x$.

In this example, the complete propagator can be obtained exactly. First, the stochastic chain rule~\cite{book:oksendal} yields $\d\ln X_t=(1/X_t)\d X_t+(\sigma^2 X_t^2/2)(-1/X_t^2)\d t$. Using  Eq.~\eqref{eq:SDE_GBM}, we have $\d\ln X_t=(\mu-\sigma^2/2)\d t+\sigma \d W_t$. Integrating from $t$ to $t+\Delta t$ gives the explicit solution of the SDE for any time duration $\Delta t$ (short or long),
\begin{equation}
    \ln X_{t+\Delta t}-\ln X_t=\left(\mu-\frac{\sigma^2}{2}\right)\Delta t+\sigma\Delta W_t\,.
\end{equation}
Much like in Sec.~\ref{sec:propSIMP}, we have written the stochastic process $X_t$ in terms of $\Delta W_t$ only. Using the normal distribution of $\Delta W_t$ and the Jacobian $\partial\Delta W_t/\partial X_t=1/(\sigma X_t)$, we find the exact probability to jump from $X_t=x$ to $X_{t+\Delta t}=x+\Delta x$ after arbitrary time $\Delta t$~\cite{book:oksendal},
\begin{equation}
    P(x+\Delta x,t+\Delta t|x,t)=\frac{1}{\sqrt{2\pi\sigma^2\Delta t}}\frac{1}{x+\Delta x}\exp\left\{-\frac{[\ln(x+\Delta x)-\ln x-(\mu-\sigma^2/2)\Delta t]^2}{2\sigma^2\Delta t}\right\}\,.\label{eq:propGBM}
\end{equation}
Note that the exact propagator is not Gaussian.

Equipped with the exact propagator, we can expand it directly up to order $\Delta t$. The leading order reads,
\begin{equation}
    P_{1/2}(x+\Delta x,t+\Delta t|x,t)=\frac{1}{\sqrt{2\pi\sigma^2\Delta t}}\frac{1}{x}\exp\left[-\frac{(\Delta x-\mu x\Delta t)^2}{2\sigma^2x^2\Delta t}\right]\,.
\end{equation}
It is Gaussian in $\Delta x$ and has standard deviation of order $\Delta t^{1/2}$. Hence, it corresponds to an $\calo (\Delta t^{1/2})$ displacement.
We proceed to obtain the first two corrections to this expression which we derived in general in the main text. Expanding formally in $\Delta x$, $1/(1+\Delta x/x)=1-\Delta x/x+\Delta x^2/x^2+\calo(\Delta t^{3/2})$ and $\ln(1+\Delta x/x)=\Delta x/x-\Delta x^2/(2x^2)+\Delta x^3/(3x^3)+\calo(\Delta t^2)$, we find
\begin{eqnarray}
   \nonumber\frac{P(x+\Delta x,t+\Delta t|x,t)}{P_{1/2}(x+\Delta x,t+\Delta t|x,t)}&=&1+\frac{1}{2\sigma^{2}x^{3}}\left(\frac{\Delta x^{3}}{\Delta t}-3\sigma^{2}x^{2}\Delta x\right)-\frac{\mu-\sigma^2/2}{2\sigma^{2}x^{2}}\left(\Delta x^{2}-\sigma^{2}x^{2}\Delta t\right)\\
   \nonumber&+&\frac{2}{3\sigma^{2}x^{4}}\left(\frac{\Delta x^{4}}{\Delta t}-6\sigma^{2}x^{2}\Delta x^{2}+3\sigma^{4}x^{4}\Delta t\right) \\
   &+&\frac{1}{8\sigma^{4}x^{6}}\left(\frac{\Delta x^{6}}{\Delta t^{2}}-15\sigma^{2}x^{2}\frac{\Delta x^{4}}{\Delta t}+45\sigma^{4}x^{4}\Delta x^{2}-15\sigma^{6}x^{6}\Delta t\right)
   +\calo(\Delta t^{3/2})\,.\label{eq:prop3/2GBM}
\end{eqnarray}
where we have written out explicitly the Hermite polynomials. Indeed, this coincides with our order-$\Delta t^{3/2}$ propagator, Eq.~\eqref{eq:prop3/2}, upon inserting $a(x)=\mu x$ and $D(x)=\sigma^2x^2/2$ in Eqs.~\eqref{eq:psi2}, \eqref{eq:psi4}, and~\eqref{eq:psi6}. Note also that the order-$\Delta t^{1/2}$ term is exactly the correction that appears in the Stratonovich ($\alpha=1/2$) version of the Euler-Maruyama propagator, Eq.~\eqref{eq:P^a_dt_incomp}. This is expected as in one dimension, the Stratonovich convention coincides with the Milstein propagator. Equation~\eqref{eq:prop3/2GBM} can be brought to the general form, Eq.~\eqref{eq:propSTRUC}, with the explicit expressions for the polynomials $\Phi$ and $\Psi$ of $y=\Delta x^2/\Delta t$, 
\begin{eqnarray}
    P(x+\Delta x,t+\Delta t|x,t) &=& P_{1/2}(x+\Delta x,t+\Delta t|x,t)\left[1+\Delta x \Phi \left(\left.\frac{\Delta x^2}{\Delta t}\right|x,t\right)+\Delta t\Psi\left(\left.\frac{\Delta x^2}{\Delta t}\right|x,t\right)+\calo(\Delta t^{3/2})\right]\,,\, \\
    \Phi (y|x,t) &=& \frac{1}{2 \sigma^2 x^3} y - \frac{3}{2x}\,, \\
    \Psi (y|x,t) &=& \frac{1}{8 \sigma^4 x^6} y^3 - \frac{29}{24 \sigma^2 x^4} y^2 + \frac12\left( \frac{15}{4 x^2} - \mu \right) y + \frac12\left( \mu - \frac{\sigma^2}{4} \right)\,.
\end{eqnarray}

With the exact propagator, Eq.~\eqref{eq:propGBM}, we can compute various functions, including the log-ratio of forward and backward propagators for arbitrary $\Delta t$. Dividing by $\Delta t$ and taking the limit $\Delta t\to0$ gives the informatic heat rate. In GBM, the drift and diffusivity are time (protocol) independent, so
\begin{equation}
    P(x,t|x+\Delta x,t+\Delta t)=\frac{1}{\sqrt{2\pi\sigma^2\Delta t}}\frac{1}{x}\exp\left\{-\frac{[\ln x-\ln(x+\Delta x)-(\mu-\sigma^2/2)\Delta t]^2}{2\sigma^2\Delta t}\right\} \, .
\end{equation}
Thus, we have
\begin{equation}
    \ln\frac{P(X_t+\Delta X_t,t+\Delta t|X_t, t)}{P(X_t,t|X_t+\Delta X_t,t+\Delta t)}=\frac{2(\mu-\sigma^2)}{\sigma^2}\ln\left(1+\frac{\Delta X_t}{X_t}\right)\,.
\end{equation}
Upon expanding $\ln(1+\Delta X_t/X_t)=\Delta X_t/X_t-\Delta X_t^2/(2X_t^2)+\calo(\Delta t^{3/2})= \dot X_t\Delta t/X_t-(\sigma X_t)^2\Delta t/(2X_t^2)+\calo(\Delta t^{3/2})$, we find by definition, Eq.~\eqref{eq:infoheat},
\begin{equation}
    \dot\Omega_t=\frac{2(\mu-\sigma^2)}{\sigma^2X_t}\dot X_t-\mu+\sigma^2\,.
\end{equation}
This coincides with Eq.~\eqref{eq:infoheatRES}, $\d\Omega_t=\{[a(X_t)-D'(X_t)]/D(X_t)\}\cdot\d X_t+\{[a(X_t)-D'(X_t)]/D(X_t)\}'D(X_t)\d t$, where $a(x)=\mu x$, $D(x)=\sigma^2x^2/2$, and $[\cdot]'=(d/dx)[\cdot]$.

\section{Non-It\^o Stochastic Taylor expansions}\label{sec:nonitostoc}

While Eq.~\eqref{eq:dXt_alpha_apprx} may seem like a natural extension of the Euler-Maruyama method to arbitrary convention $\alpha$, the mathematically-consistent machinery for obtaining short-time expansions is rather different (see Sec.~\ref{sec:expans}, based upon Ref.~\cite{book:KloedenPlaten}). Evaluating the amplitude $\vecb$ at the intermediate  point introduces a shift in the higher-order terms and the `overshot' drift $\veca$.
Particularly, as seen below, Eq.~\eqref{eq:dXt_alpha_apprx} is also only correct to order-$\Delta t^{1/2}$, meaning that it and Eq.~\eqref{eq:dXnaive} coincide, and there indeed is no convention dependence.
We proceed to show that, in general, there is no advantage in using non-It\^o conventions, especially for cases where the leading-order propagator sufficies.

To benchmark the upcoming comparison, we recall the non-It\^o Milstein scheme. We follow the steps presented in Sec.~\ref{sec:expans}, but now for Eq.~\eqref{eq:langevin_alpha} with the general convention $\alpha$. The differences between Eqs.~\eqref{eq:dX1alpha} and~\eqref{eq:dXt_alpha_apprx} will already appear within the order-$\Delta t$ term. Expanding Eq.~\eqref{eq:langevin_alpha} to order $\Delta t$,
\begin{equation}
    \Delta X^\mu_t=a_\alpha^\mu(\vecX_{t},\Lambda_{t})\Delta t + b^{\mu\nu}(\vecX_{t},\Lambda_{t})\Delta W^\nu_{t}+  [b^{\sigma\rho}\nabla^\sigma b^{\mu\nu}](\vecX_{t},\Lambda_{t})\int_{t}^{t+\Delta t}\d W^\nu_{t'}\acirc\int_{t}^{t'}\d W^\rho_{t''}+\calo(\Delta t^{3/2})\,.\label{eq:dX1alpha}
\end{equation}
Notice that now, each time the differential $\d\vecW_t$ appears, it multiplies the integrands as $\acirc$. For example, trivially, $\int_t^{t+\Delta t}\d\vecW_t\acirc1=\Delta\vecW_t$. To compare Eq.~\eqref{eq:dX1alpha} to the It\^o Milstein scheme (Eq.~\eqref{eq:dX1}), we transition from the double stochastic integral within the $\alpha$-convention to the It\^o one. We finely-discretize as before, and obtain
\begin{eqnarray}
    \int_{t}^{t+\Delta t}\d W^\nu_{t'}\acirc\int_{t}^{t'}\d W^\rho_{t''}
    &=&\lim_{\delta t\to0}\sum_{i=1}^{\Delta t/\delta t}(W^\nu_{t+i\delta t}-W^\nu_{t+(i-1)\delta t})[W^\rho_{t+(i-1)\delta t}+\alpha(W^\nu_{t+i\delta t}-W^\nu_{t+(i-1)\delta t})-W^\rho_t] \nonumber\\
    &=&\lim_{\delta t\to0}\sum_{i=1}^{\Delta t/\delta t}(W^\nu_{t+i\delta t}-W^\nu_{t+(i-1)\delta t})(W^\rho_{t+(i-1)\delta t}-W^\rho_t)\nonumber\\
    &+&\lim_{\delta t\to0}\sum_{i=1}^{\Delta t/\delta t}\alpha(W^\nu_{t+i\delta t}-W^\nu_{t+(i-1)\delta t})(W^\rho_{t+i\delta t}-W^\rho_{t+(i-1)\delta t})\nonumber\\
    &=&\int_{t}^{t+\Delta t}\d W^\nu_{t'}\cdot\int_{t}^{t'}\d W^\rho_{t''}+\alpha\delta^{\nu\rho}\int_t^{t+\Delta t}\d t'\,.
\end{eqnarray}
Using $a^\mu_\alpha+\alpha b^{\sigma\rho}\nabla^\sigma b^{\mu\nu}\delta^{\nu\rho}=a^\mu$, we notice that Eq.~\eqref{eq:dX1alpha} coincides \emph{exactly} with Eq.~\eqref{eq:dX1} to order $\Delta t$, as it must. This suggests that both expansion were done consistently, and all relevant terms were included. Indeed, again there is no dependence on convention and discretization. 

To contrast the generalized Euler-Maruyama approximation and the rigorous stochastic Taylor expansion, we bring Eq.~\eqref{eq:dXt_alpha_apprx} to It\^o form as well, Eq.~\eqref{eq:dX1/2}, by once again expanding the coefficients around $\vecX_t$ and keeping order-$\Delta t$ terms:
\begin{eqnarray}
\nonumber    \Delta X^\mu_t
&\simeq&a_\alpha^\mu(\vecX_{t},\Lambda_{t})\Delta t + b^{\mu\nu}(\vecX_{t},\Lambda_{t})\Delta W^\nu_{t}+  \alpha\nabla^\sigma b^{\mu\nu}(\vecX_{t},\Lambda_{t})\Delta W^\nu_{t}\Delta X^\sigma_{t}+\calo(\Delta t^{3/2})\\
    &\simeq&a^\mu(\vecX_{t},\Lambda_{t})\Delta t + b^{\mu\nu}(\vecX_{t},\Lambda_{t})\Delta W^\nu_{t}+  \alpha[b^{\sigma\rho}\nabla^\sigma b^{\mu\nu}](\vecX_{t},\Lambda_{t})(\Delta W^\nu_{t}\Delta W^\rho_{t}-\delta^{\nu\rho}\Delta t)+\calo(\Delta t^{3/2})\,.\label{eq:dXt_alpha_expand}
\end{eqnarray}
In this expression, we see that the first two terms agree with the rigorous Euler-Maruyama method expansion, Eq.~\eqref{eq:dX1/2}. However, the third term of order $\Delta t$, $\alpha[\Delta W^\nu_{t}\Delta W^\rho_{t}-\delta^{\nu\rho}\Delta t]$, differs from the third term in the rigorous and coinciding Eqs.~\eqref{eq:dX1} and~\eqref{eq:dX1alpha}, $\int_{t}^{t+\Delta t}\d W^\nu_{t'}\cdot\int_{t}^{t'}\d W^\rho_{t''}$.
For example, while both have mean $0$, their variances are different: $\alpha^2\langle[\Delta W^\mu_{t}\Delta W^\sigma_{t}-\delta^{\mu\sigma}\Delta t][\Delta W^\nu_{t}\Delta W^\rho_{t}-\delta^{\nu\rho}\Delta t]\rangle=(\alpha^2\Delta t^2)[\delta^{\mu\nu}\delta^{\sigma\rho}+\delta^{\mu\rho}\delta^{\sigma\nu}]$  while $\langle[\int_{t}^{t+\Delta t}\d W^\mu_{t'}\cdot\int_{t}^{t'}\d W^\sigma_{t''}][\int_{t}^{t+\Delta t}\d W^\nu_{s'}\cdot\int_{t}^{s'}\d W^\rho_{s''}]\rangle=(\Delta t^2/2)\delta^{\mu\nu}\delta^{\sigma\rho}$. 
These two terms happen to agree only in one dimension for $\alpha=1/2$ (Stratonovich), since then, using the It\^o lemma~\cite{book:oksendal},
\begin{equation}
    \int_{t}^{t+\Delta t}\d W^\nu_{t'}\cdot\int_{t}^{t'}\d W^\rho_{t''}=\int_{t}^{t+\Delta t}\d W^\nu_{t'}\circ(W_{t'}-W_t)-\frac12\int_{t}^{t+\Delta t}\d t'=\frac12[\Delta W_t^2-\Delta t]\,.
\end{equation}

Therefore, the generalized approximation adopted in the literature~\cite{LubenskyPRE07}, Eq.~\eqref{eq:dXt_alpha_apprx}, is no more useful than the standard Euler-Maruyama method, Eq.~\eqref{eq:dX1/2}, except in the very particular case of one-dimensional Stratonovich convention. In  Ref.~\cite{DrozdovPRE97}, the early truncation of the next-leading-order expansion of the propagator's moment generating function has indeed yielded a Stratonovich-like propagator. This also ignores the additional terms in the order-$\Delta t^{3/2}$ expansions, Eq.~\eqref{eq:dX3/2}, that may be needed for various other functionals of the propagator. A similar order-$\Delta t^{3/2}$ expansion as Eq.~\eqref{eq:dX1alpha} for a non-It\^o convention can be found in the literature~\cite{book:KloedenPlaten}.

With that, we understand why, for the specific case of $\dot\Omega_t$, propagators of ``complementary'' conventions are sufficient: First, the order-$\Delta t$ terms (coming from the order-$\Delta t^{3/2}$ expansion) vanish as seen in Eq.~\eqref{eq:EPRcorr}. Second, the order-$\Delta t^{1/2}$ terms (coming from the Milstein expansion) take the form $\int_t^{t+\Delta t}\d\vecW_{t'}\int_t^{t'}\d\vecW_{t''}$. For the reverse propagator, we should have computed $\int_{t+\Delta t}^t\d\vecW_{t'}\int_{t+\Delta t}^{t'}\d\vecW_{t''}$.
Summing the two gives $\int_t^{t+\Delta t}\d\vecW_{t'}\int_t^{t'}\d\vecW_{t''}+\int_{t+\Delta t}^t\d\vecW_{t'}\int_{t+\Delta t}^{t'}\d\vecW_{t''}=\Delta\vecW_t\Delta\vecW_t-\int_t^{t+\Delta t}\vecD(\vecX_{t'},t')\d t'$, where the second term is a consequence of the It\^o product within the double stochastic integral.
If, on the other hand, we choose to use the complementary conventions from Eq.~\eqref{eq:dXt_alpha_apprx}, in light of Eq.~\eqref{eq:dXt_alpha_expand}, we would have  $\alpha[\Delta\vecx\Delta\vecx-D(\vecx+\alpha\Delta\vecx,\lambda)]$ and $(1-\alpha)[\Delta\vecx\Delta\vecx-D(\vecx+(1-\alpha)\Delta\vecx,\lambda)]$ for the forward and backward propagators. Their sum, to order $\Delta t$, is also $\Delta\vecW_t\Delta\vecW_t-\vecD(\vecX_{t+\Delta t/2},t)\Delta t$, which agrees with the exact results up to order $\Delta t$. Thus, we have two separate cancellation of errors, specific for the informatic heat, that allows using lower-order propagators than necessary.
It is important to stress that, although intuitive, it is impossible to deduce this cancellation without actually performing the expansion.

\section{Moments of all Euler-Maruyama propagators are correct to the same order}\label{sec:moment}

For completeness, we show here how the it\^o Euler-Maruyama expansion, Eq.~\eqref{eq:dX1/2}, suffices for computing moments to leading order, and that its extension to all $\alpha$ conventions, Eq.~\eqref{eq:dXt_alpha_apprx}, gives the same results.
Since there is only one Fokker-Planck equation and thus a single stochastic differential equation (see sec.~\ref{sec:dilemma}), the propagators must be identical for all $\alpha$ at any order. 
Thus, upon expanding $\veca_\alpha(\vecx+\alpha\Delta\vecx)$ and $\vecb(\vecx+\alpha\Delta\vecx)$ (as appearing in $P_{1/2}^\alpha$) around $\vecx$ must give back $P_{1/2}^0\equiv P_{1/2}$, Eq.~\eqref{eq:propSIMP}, to order $\Delta t$. 

To this end, one should write the expansion $[\cdot](\vecx+\alpha\Delta\vecx)=[\cdot](\vecx)+\alpha\Delta\vecx\cdot\grad[\cdot](\vecx)+(\alpha^2/2)\Delta\vecx\Delta\vecx:\grad\grad[\cdot](\vecx)$ for $a^\mu_\alpha=a^\mu_0-\alpha b^{\sigma\nu}\nabla^\sigma b^{\mu\nu}$ and $\vecb$, and keep terms to order $\Delta t$. For the ease of demonstration, we show the result for the simple $d=1$ case,
\begin{eqnarray}
    P_{1/2}^\alpha(x+\Delta x,t+\Delta t|x,t)&=&\frac{1}{\sqrt{4\pi D(x+\alpha\Delta x)\Delta t}}\times\nonumber\\
    &\times&\exp\left\{-\frac{[\Delta x-a_\alpha(x+\alpha\Delta x)\Delta t+\alpha D'(x+\alpha\Delta x)\Delta t]^2}{4D(x+\alpha\Delta x)\Delta t}-\alpha a_\alpha'(x+\alpha\Delta x)\Delta t\right\} \nonumber\\
    &=&P_{1/2}^0(x+\Delta x,t+\Delta t|x,t)\left[1-\frac{\alpha D'(x)}{4D^2(x)}\left(6D(x)\Delta x-\frac{\Delta x^3}{\Delta t}\right)+\calo(\Delta t^{3/2})\right]\,,\label{eq:P^a_dt_incomp}
\end{eqnarray}
which required using $\Delta x^2=2D(x)\Delta t+\calo(\Delta t^{3/2})$, $\Delta x^4/\Delta t=12D(x)\Delta t+\calo(\Delta t^{3/2})$, and $\Delta x^6/\Delta t^2=120D(x)\Delta t+\calo(\Delta t^{3/2})$, and we abbreviated $\partial[\cdot]/\partial x=[\cdot]'$. 
Observe the consistency of the expansions performed in this paper: In these one-dimensional expressions, only the Stratonovich propagator [$\alpha=1/2$ in Eq.~\eqref{eq:P^a_dt_incomp}] agrees with the Milstein propagator, Eq.~\eqref{eq:prop1}, as we predicted in Appendix~\ref{sec:nonitostoc}.

Another verification we can make is that the normalization and the first two conditional moments using the $\alpha$ propagator are indeed correct \emph{and $\alpha$-independent} to the appropriate order,
\begin{eqnarray}
    \langle{\Delta x^n}|x\rangle_\alpha&\equiv&\int \d\Delta x P_{1/2}^\alpha(x+\Delta x,t+\Delta t|x,t)\Delta x^n \nonumber\\
    &=&\langle\Delta x^n|x\rangle_0-\frac{\alpha D'(x)}{4D^2(x)}\left(6D(x)\langle\Delta x^{n+1}|x\rangle_0-\frac{\langle\Delta x^{n+3}|x\rangle_0}{\Delta t}\right)+\calo(\Delta t^2) \nonumber\\
    &=&\begin{cases}
1+\calo(\Delta t^{2}), & n=0\,,\\
a(x)\Delta t+\calo(\Delta t^{2}), & n=1\,,\\
2D(x)\Delta t+\calo(\Delta t^{2}), & n=2\,,\\
\calo(\Delta t^{2}), & n\geq3\,,
\end{cases}
\end{eqnarray}
where we used $\langle\Delta x|x\rangle_0=a(x)\Delta t$, $\langle\Delta x^3|x\rangle_0/\Delta t=6D(x)a(x)\Delta t$, and $\langle\Delta x^{n\geq3}|x\rangle_0,\langle\Delta x^{n\geq5}|x\rangle_0/\Delta t\sim\Delta t^2$. Therefore, the moments and the Fokker-Planck equation (given by its Kramers-Moyal expansion~\cite{book:FPE}), as calculated for example in~\cite{book:schuss}, are accurately reproduced from all Euler-Maruyama methods, Eqs.~\eqref{eq:dX1/2} and~\eqref{eq:dXt_alpha_apprx}.

We conclude that using the $\alpha$ convention, Eq.~\eqref{eq:prop_alpha}, does not provide any improved accuracy compared to the usual It\^o Euler-Maruyama propagator, Eq.~\eqref{eq:prop1/2}. If one observes a dependence on discretization while using different choices of $\alpha$ in Eq.~\eqref{eq:prop_alpha}, it means that a higher-order propagator is needed, \eg Eq.~\eqref{eq:prop1} or even Eq.~\eqref{eq:prop3/2}.

\end{document}